\documentclass[reprint, aps]{revtex4-2}
\usepackage{amsmath,amssymb,mathrsfs,multirow}
\usepackage{graphicx}
\usepackage{dcolumn}
\usepackage{bm}
\usepackage[colorlinks=true,linkcolor=blue,urlcolor=blue,citecolor=blue]{hyperref}

\allowdisplaybreaks
\begin{document}

\title{Optimal estimation of Higgs-Gauge Boson couplings at the future $e^+e^-$ colliders}

\author{Subhaditya Bhattacharya}
\email{subhab@iitg.ac.in}
\author{Amir Subba}
\email{amirsubba@iitg.ac.in}
\author{Abhik Sarkar}
\email{sarkar.abhik@iitg.ac.in}
\affiliation{Department of Physics, Indian Institute of Technology Guwahati, Guwahati, Assam, 781039, India}

\begin{abstract}
 The proposed $e^+e^-$ collider offers an ideal environment for precise estimation of Higgs boson properties which are of utmost importance to validate the 
 Standard Model of particle physics. We investigate $hVV$ couplings, where $V\in \{Z,\gamma\}$ with single Higgs production associated with $Z$ boson at the proposed 
 $e^+e^-$ machine with $\sqrt{s}=250$ GeV, within the Standard Model Effective Field Theory (SMEFT) framework. 
 We employ the recoil mass of the dilepton system, to select the signal phase space, i.e, $Zh \to l^+l^-b\Bar{b}$ events. 
 The constraints on the Wilson coefficients (WCs) are obtained using the optimal observable technique (OOT). 
 On comparison with the current experimental limits at $68\%$ CL with $138$ fb$^{-1}$ luminosity, our limits are tighter by a factor ranging from $1.5-10$ for CP even operators, while CP-odd WCs shows comparable limits. 
\end{abstract}

\maketitle
\section{Introduction}
\label{sec:intro}
The discovery of the Higgs boson at the Large Hadron Collider (LHC) by the ATLAS~\cite{ATLAS:2012yve} and CMS~\cite{CMS:2012qbp,CMS:2013btf} collaborations unleashes
 a cornerstone achievement in particle physics, completing the particle spectrum of the Standard Model~(SM)~\cite{Weinberg:1967tq}. The Higgs field endows the 
$W^\pm$ and $Z$ bosons, as well as the SM fermions with mass through electroweak symmetry breaking (EWSB)~\cite{Englert:1964et,Guralnik:1964eu,Higgs:1964pj}, 
while preserving the $SU(3)_c\times SU(2)_L \times U(1)_Y$ gauge invariance and renormalizability. 
Although the SM provides a remarkably successful theory up to the electroweak scale and has passed through numerous precision tests, it leaves 
several fundamental questions unresolved. Among these are the naturalness of the electroweak scale, the origin and structure of the Yukawa couplings 
responsible for fermion mass hierarchies, the mechanism of underlying CP violation, the origin of neutrino masses and mixing etc. 
In the SM framework, such features are often accommodated through ad hoc parameters without offering  a dynamical explanations. Furthermore, 
the SM does not contain a viable dark matter candidate and fails to account for the observed baryon asymmetry of the Universe. Given that the Higgs boson 
is intimately connected to the mechanism of mass generation and the vacuum structure, it is widely anticipated that it may serve as a portal to beyond-the-SM (BSM) physics.

A multitude of experimental investigations at LHC~\cite{CMS:2012vby,CMS:2013fjq,CMS:2014nkk,CMS:2015chx,CMS:2016tad,CMS:2017len,CMS:2019ekd,CMS:2019jdw,ATLAS:2013xga,ATLAS:2015zhl,ATLAS:2016ifi,ATLAS:2017azn,ATLAS:2017qey,ATLAS:2018hxb,ATLAS:2020evk} have imposed stringent constraints on the spin-parity quantum numbers of the Higgs boson, as well as its interactions with gluons and electroweak (EW) gauge bosons. Current measurements are in agreement with the SM prediction of a scalar boson with quantum numbers $J^{PC}=0^{++}$. Nonetheless, the presence of small deviations from the SM in the form of anomalous couplings remains experimentally permissible. In BSM theories, interactions with the Higgs boson may occur through several anomalous couplings, which lead to new tensor structures in the interaction terms that can be both CP-even or CP-odd. The hadronic environment of the LHC imposes limitations on measurement precision due to QCD backgrounds, pile-up, and uncertainties associated with parton distribution functions (PDFs). This has motivated the proposal of high-luminosity and high-precision lepton colliders, such as the ILC~\cite{ILC:2007bjz,ILC:2007oiw,ILC:2007vrf,ILC:2013jhg,Adolphsen:2013kya}, FCC-ee~\cite{FCC:2018evy}, CEPC~\cite{CEPCStudyGroup:2018rmc,CEPCStudyGroup:2023quu,CEPCPhysicsStudyGroup:2022uwl} and CLIC~\cite{CLIC:2016zwp,Aicheler:2012bya,Linssen:2012hp,Lebrun:2012hj}, which offer a complementary environment for Higgs boson studies. 
Projected experimental uncertainties on Higgs couplings at future $e^+e^-$ colliders are expected to reach the per-mille level~\cite{ILC:2013jhg,ILCInternationalDevelopmentTeam:2022izu,LinearCollider:2025lya,LinearColliderVision:2025hlt,Roloff:2018dqu}, 
significantly surpassing the capabilities of the LHC.

A particularly clean setting for precision Higgs physics is provided by an $e^+e^-$ collider operating at a center-of-mass energy of $\sqrt{s} = 250$~GeV, which is the focus of current work. In this energy regime, the dominant Higgs production channel is the Higgs-strahlung process, $ e^+e^- \rightarrow Zh$, which proceeds via $s$-channel exchange of a $Z$ boson. The Higgs-strahlung production mechanism allows for model-independent measurements of Higgs boson properties by reconstructing the Higgs mass from the recoil mass spectrum of the $Z$ boson, independent of the Higgs decay mode. Moreover, the ability to polarize the initial $e^-$ and $e^+$ beams in the linear collider provides an additional handle to disentangle helicity structures of electroweak interactions and enhance sensitivity to specific interaction channels.

In the current work, we study the potential of ILC to probe the anomalous Higgs-gauge couplings in $Zh$ production process within the SM effective field theory (SMEFT) framework. The absence of direct evidence for new states at the LHC suggests that new degrees of freedom may reside at energy scales higher than currently accessible, with their low-energy imprints manifesting as subtle modifications to SM couplings. In such scenario, SMEFT provides a powerful tool to parameterize deviations from the SM predictions in a systematic and gauge-invariant manner, wherein higher-dimensional operators are added to the SM Lagrangian. Our aim is to obtain the optimal limit on the operator coefficients 
contributing to $hZZ$ vertex and $Zh$ production at ILC.

The current work is organized as follows: In Sec.~\ref{sec:SMEFT}, we discuss the SMEFT framework and the relevant operators. In 
Sec.~\ref{sec:ml}, we discuss the event selection using the dilepton recoil mass observable. We focus on $l^+l^-b\Bar{b}$ final state owing to large Higgs branching fraction to $b\Bar{b}$ channel. In Sec.~\ref{sec:oot}, we discuss in brief the optimal observable technique. In Sec.~\ref{sec:result}, we discuss the one parameter limits on Wilson coefficients obtained for unpolarized and polarized beams scenario. We conclude in Sec.~\ref{sec:conclude}.

\section{SMEFT framework and relevant operators}
\label{sec:SMEFT}

The SMEFT Lagrangian is written as a perturbative expansion in terms of the inverse power of  NP scale $\Lambda$ where decoupling occurs, accompanied with 
$\mathcal{O}_i$ operators of dimension-$d$ constructed of the SM fields obeying SM gauge invariance \cite{Buchmuller:1985jz},
\begin{equation}
    \mathscr{L}_{\text{SMEFT}} = \mathscr{L}_{\text{SM}} + \sum_i \frac{C_i^{(5)}}{\Lambda}\mathscr{O}_i^{(5)} + \sum_{i} \frac{C_i^{(6)}}{\Lambda^2} \mathscr{O}_i^{(6)} + \mathscr{O}\left(\frac{1}{\Lambda^4}\right),
\end{equation}
where $C_i$ are the Wilson coefficients. Since at each order of $d$, the amplitude becomes suppressed as $(E/\Lambda)^{(d-4)}$, 
the dominant SMEFT contribution comes from lowest dimension operator that contributes to the process. Also the SMEFT construction remains valid only 
when the scale of the reaction remains below the cut off scale $\Lambda$, requiring us to adhere to CM energy of the collider $\sqrt{s}<\Lambda$. Throughout the analysis, we set $\Lambda = 1$ TeV. Assuming the lepton and baryon number conservation, the lowest order of SMEFT begins at dimesion-$6$. We consider three CP-even and 
their dual CP-odd operators inducing anomalous contribution to $Zh$ process. The relevant operators in Warsaw basis are~\cite{Grzadkowski:2010es},
    \begin{align}
        \label{eq:dim6}
            \mathscr{O}_{H W} = (H^{\dagger} H) W^{i}_{\mu \nu} W^{i \mu \nu},\hspace{0.8cm}\mathscr{O}_{H\widetilde{W}} &= (H^{\dagger} H) \widetilde{W}^{i}_{\mu \nu} W^{i \mu \nu} \nonumber\\
            \mathscr{O}_{H WB} = (H^{\dagger} \tau^{i} H) W^{i}_{\mu \nu} B^{\mu \nu},\hspace{0.4cm} \mathscr{O}_{H\widetilde{W}B} &= (H^{\dagger} \tau^{i} H) \widetilde{W}^{i}_{\mu \nu} B^{\mu \nu}\nonumber \\
            \mathscr{O}_{H B} = (H^{\dagger} H) B_{\mu \nu} B^{\mu \nu}, \hspace{1.0cm}\mathscr{O}_{H\widetilde{B}} &= (H^{\dagger} H) \widetilde{B}_{\mu \nu} B^{\mu \nu}\,.
    \end{align}
Here, the operators $\mathscr{O} \in \{\mathscr{O}_{H W},\mathscr{O}_{H B},\mathscr{O}_{H WB}\}$ are CP-even while the remaining subset of operators are CP-odd. The field tensor are defined as $W_{\mu \nu}^{i} = \partial_{\mu} W_{\nu}^{i} - \partial_{\nu} W_{\mu}^{i} + g \epsilon^{ijk} W_{\mu}^{j} W_{\nu}^{k}$, $B_{\mu \nu} = \partial_{\mu} B_{\nu} - \partial_{\nu} B_{\mu}$ and the dual field tensor is $\widetilde{V}_{\mu \nu} = \epsilon_{\mu \nu \rho \sigma} V^{\rho \sigma}$ ($V = W^{i}, B$), with the Levi-Civita 
tensor $\epsilon_{\mu \nu \rho \sigma}$; following standard notation $\epsilon_{0123}=1$. $H$ denotes the SM Higgs isodoublet. Upon spontaneous symmetry 
breaking it acquires non-zero vacuum expectation value $v$, which takes the form $\langle H\rangle=\frac{1}{\sqrt{2}}\begin{pmatrix} 0\\ v+h \end{pmatrix}$ in unitary gauge, where 
$h$ denotes SM Higgs field. These operators thereafter induces anomalous contribution to $hZZ$ coupling of the form
\begin{equation}
\begin{split}
    \delta\, \Gamma_{hZZ} &= \kappa_{hZZ} \left(\frac{h}{v} Z_{\mu \nu} Z^{\mu \nu} \right) + \kappa_{h\widetilde{Z}Z} \left(\frac{h}{v} \widetilde{Z}_{\mu \nu} Z^{\mu \nu} \right),
\end{split}
\end{equation}
where
\begin{align}
    \kappa_{hZZ} &= \frac{2v^2}{\Lambda^2} \left[\cos^{2}{\theta_{W}}\,C_{H W} + \cos{\theta_{W}}\sin{\theta_{W}}\,C_{H WB} \right.\nonumber\\&+\left. \sin^{2}{\theta_{W}}\,C_{H B}\right],\nonumber\\
    \kappa_{h\widetilde{Z}Z} &= \frac{2v^2}{\Lambda^2} \left[\cos^{2}{\theta_{W}}\,C_{H \widetilde{W}} + \cos{\theta_{W}}\sin{\theta_{W}}\,C_{H \widetilde{W}B} \right.\nonumber\\&+\left. \sin^{2}{\theta_{W}}\,C_{H \widetilde{B}}\right].
\end{align}
 Apart from anomalous $hZZ$ coupling, the operators allows for tree level $s$-channel $Zh$ production mediated by massless photon, which otherwise is a loop process in the SM. The anomalous $hZ\gamma$ coupling is obtained as,
\begin{equation}
    \delta\, \Gamma_{hZ\gamma} = \kappa_{hZ\gamma} \left(\frac{h}{v} Z_{\mu \nu} A^{\mu \nu} \right) + \kappa_{h\widetilde{Z}\gamma} \left(\frac{h}{v} \widetilde{Z}_{\mu \nu} A^{\mu \nu} \right),
\end{equation}
where
    \begin{align}
    \kappa_{hZ\gamma} &= \frac{v^2}{\Lambda^2} \left[2\cos{\theta_{W}}\sin{\theta_{W}}\,(C_{H W} - C_{H B}) \right.\nonumber\\&+\left. (\sin^{2}{\theta_{W}} - \cos^{2}{\theta_{W}})\,C_{H WB}\right],\nonumber\\
    \kappa_{h\widetilde{Z}\gamma} &= \frac{v^2}{\Lambda^2} \left[2\cos{\theta_{W}}\sin{\theta_{W}}\,(C_{H \widetilde{W}} - C_{H \widetilde{B}}) \right.\nonumber\\&+\left. (\sin^{2}{\theta_{W}} - \cos^{2}{\theta_{W}})\,C_{H \widetilde{W}B}\right].
    \end{align}
We do not consider operators affecting $f\Bar{f}V$ or $f\Bar{f}VH$ couplings as they are tightly constrained from LEP data. Various studies~\cite{Hagiwara:1993sw,Hagiwara:2000tk,Biswal:2005fh,Biswal:2008tg,Biswal:2009ar,Godbole:2007cn,Godbole:2013saa,Godbole:2014cfa,Rindani:2009pb,Rindani:2010pi,Anderson:2013afp,Craig:2015wwr,Beneke:2014sba,Khanpour:2017cfq,Zagoskin:2018wdo,Li:2019evl,He:2019kgh,Nakamura:2017ihk,Banerjee:2019pks,Chiu:2017yrx,Durieux:2017rsg,Rao:2020hel,Bizon:2021rww,Corbett:2012dm,Corbett:2012ja,Bhattacharya:2024sxl,Li:2023tcr,Xie:2021xtl,Yan:2021tmw,Cao:2018cms,Ogawa:2017bmg,Barklow:2017suo,Barklow:2017awn} has been done to probe the structure of anomalous Higgs-gauge couplings using the kinematic and angular distribution at LHC at $e^+e^-$ colliders. Quantum tomography~\cite{Fabbrichesi:2023jep,DelGratta:2025qyp,Sullivan:2024wzl,Bernal:2023ruk,Subba:2024aut} has also been used to constrain anomalous $hVV,V\in \{Z,W\}$ coupling. On the experimental side, the current limits at $68\%$ confidence level (CL) on the WCs of the above operators are provided by CMS~\cite{CMS:2024bua} (assuming $\Lambda=1$ TeV):
\begin{equation}
\begin{split}
    C_{H W} = [-0.79,+0.51], \hspace{0.6cm} C_{H \widetilde{W}} &= [-0.76,+0.41],  \\
    C_{H WB} = [-1.62,+1.50], \hspace{0.4cm} C_{H \widetilde{W}B} &= [-1.57,+0.83],\\
    C_{H B } = [-0.23,+0.16], \hspace{0.7cm}
    C_{H \widetilde{B}} &= [-0.23,+0.12].\\
\end{split}
\end{equation}

As stated before, in this work we explore the extent to which optimal observable analysis of production cross~section can constrain modifications to the Higgs-gauge couplings, 
and highlight the complementarity of this program with current LHC studies. Our analysis underscores the central role that an $e^+e^-$ Higgs factory can play in elucidating the 
nature of electroweak symmetry breaking and probing the structure of physics beyond the SM.

\begin{figure*}[htb!]
    \centering
    \includegraphics[width=0.24\linewidth]{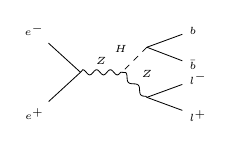}
    \includegraphics[width=0.24\linewidth]{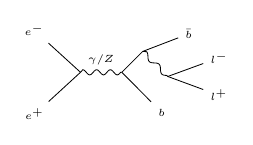}
    \includegraphics[width=0.24\linewidth]{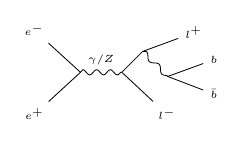}
    \includegraphics[width=0.24\linewidth]{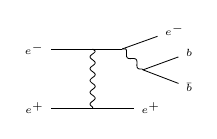}
    \includegraphics[width=0.24\linewidth]{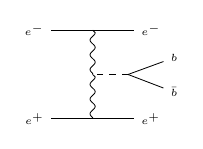}
    \includegraphics[width=0.24\linewidth]{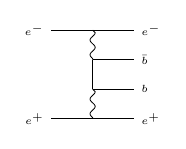}
    \includegraphics[width=0.24\linewidth]{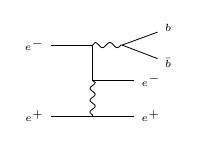}
    \includegraphics[width=0.24\linewidth]{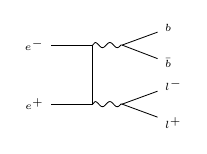}
    \caption{Representative Feynman diagrams denoting the production of two leptons and two $b$ quarks at the leading order in the SM.}
    \label{fig:feynman}
\end{figure*}

\section{Collider analysis: Event selection}
\label{sec:ml}
The signal process considered here is the Higgs-strahlung process followed by leptonic decay of $Z$ boson and Higgs boson decay to pair of $b$-quark at ILC, 
i.e., $e^+e^- \to Z(l^+l^-)h(b\overline{b})$ for the SM as well as dimension six effective operators listed in Eq.~\eqref{eq:dim6} that contribute to it. The dominant background to the $l^+l^-b\overline{b}$ final state arises from $e^+e^- \to Z(l^+l^-)Z(b\overline{b})$. A subdominant background from Z-boson fusion Higgs production, $e^+e^- \to e^+e^-h(b\overline{b})$, contributes negligibly at $\sqrt{s} = 250$~GeV and is therefore neglected. {The $ZZ\gamma$ process could also dilute the analysis, however the $p_T$ cut on final photons reduces such contribution significantly. We also consider the effect of initial state radiation (ISR) on the overall selection kinematics of the processes.} The schematic Feynman diagrams representing signal and  some of the background processes are illustrated in Fig. \ref{fig:feynman}. 

Operationally, the dim-6 operators listed in Eq.~\eqref{eq:dim6} are implemented in \texttt{FeynRules} \cite{Alloul:2013bka} to obtain an {\tt Universal FeynRules Output} (UFO) model files. The corresponding UFO model~\cite{Darme:2023jdn} is exported to \texttt{MG5\_aMC} \cite{Alwall:2011uj} for Monte Carlo event generation. The matrix level events are then passed through \texttt{Pythia8} \cite{Bierlich:2022pfr} for showering of partons and hadronization of final state particles. Finally, the detector simulation are implemented using \texttt{Delphes3} \cite{deFavereau:2013fsa} with the ILC parameters~\cite{Behnke:2013lya}. For $b$-tagging of final jets, we employ the loose working point, corresponding to an average tagging efficiency of 80\%. Event selection requires exactly two oppositely charged leptons and two $b$-tagged jets, with the additional condition that the final state contains no {hard~$(p_T > 10~\mathrm{GeV})$} photons.
\begin{figure}[htb!]
    \centering
    \includegraphics[width=0.95\linewidth]{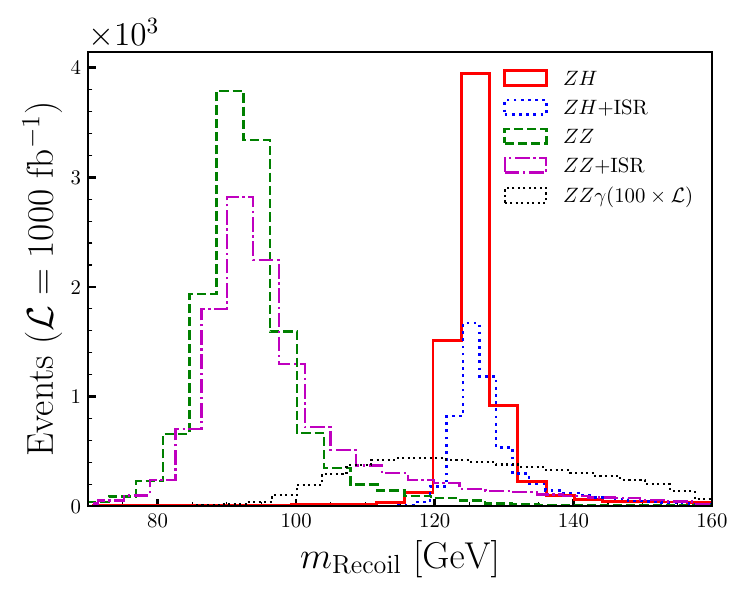}
    \caption{{Recoil mass distribution normalized to $\mathcal{L}=1000$ fb$^{-1}$ for signal and two background processes viz. $ZZ$ and $ZZ\gamma$. The kinematic effect of ISR are highlighted for $Zh$ and $ZZ$ processes.}}
    \label{fig:dist}
\end{figure}

In order to select the phase space which maximizes the significance, a number of kinematic variables are constructed using the four momenta of final four objects. We train \texttt{LightGBM} boosted decision tree (BDT) models separately for the unpolarized and polarized beam configurations: $(P_{e^+}, P_{e^-}) = (+30\%, -80\%)$ and $(-30\%, +80\%)$, respectively. The input features for the BDTs consist of several kinematic observables, whose details are provided in Appendix \ref{sec:bdt}. Among these variables, the recoil mass of the dilepton system, $m_{\rm Recoil}$, emerges as the most effective discriminator between signal and background. {The distribution normalized to $\mathcal{L}=1000$ fb$^{-1}$ representing $m_{\rm Recoil}$ for signal and two major background processes ($ZZ$ and $ZZ\gamma$) with and without ISR are shown in Fig.~\ref{fig:dist}. The $p_T$ cut in the final photons  remove the $ZZ\gamma$ significantly and the corresponding distribution are rescaled by factor of 100 in integrated luminosity for visualization. The number of events for the case of signal $(Zh)$ reduces due to ISR as the interaction energy lowers below $250$ GeV leading to lower cross~section. However, no significant change in shape and count is observed for $ZZ$ process. Nevertheless, the recoil mass stands out to nullify the effect of ISR on segregating the signal phase space from that of background.} We apply the following selection cut on the recoil mass:
\begin{align}
    |m_{\rm Recoil} - M_{h}| < 2~\mathrm{GeV}\,,
\end{align}
Events and signal significance before and after the $M_{\rm recoil}$ cut is furnished in Table~\ref{tab:cuts} for both polarized and unpolarized beams. 
After applying the recoil mass cut, we find that the background becomes nearly negligible compared to the signal.

\begin{table*}[htb!]
    \centering
    \caption{Cutflow table for signal and backgrounds, at the ILC $\sqrt{s} = 250$ GeV and an integrated luminosity of 1000 fb$^{-1}$.}
    \label{tab:cuts}
    \renewcommand{\arraystretch}{1.5}
		\begin{tabular*}{1\textwidth}{@{\extracolsep{\fill}}lccccc@{}}\hline
    \hline 
        \multirow{2}*{$(P_{e^+}, P_{e^-})$} & \multirow{2}*{Process} & \multicolumn{2}{c}{Before $M_{\rm recoil}$ cut} & \multicolumn{2}{c}{After $M_{\rm recoil}$ cut} \\ \cline{3-4} \cline{5-6}
         &  & Events & $\mathcal{Z}=S/B$ & Events & $\mathcal{Z}=S/B$ \\ \hline \hline
        \multirow{2}*{Unpolarized} & Signal (S) & 1879 & \multirow{2}*{0.11} & 1610 & \multirow{2}*{36.9} \\
         & Background (B) & 16652 & & 44 & \\ \hline
        \multirow{2}*{$(+30\%,-80\%)$} & Signal (S) & 2580 & \multirow{2}*{0.12} & 2189 & \multirow{2}*{31.2} \\
         & Background (B) & 21252 & & 70 & \\ \hline
        \multirow{2}*{$(-30\%,+80\%)$} & Signal (S) & 2045 & \multirow{2}*{0.13} & 1746 & \multirow{2}*{61.0} \\
         & Background (B) & 15722 &  & 29 & \\
         \hline \hline
    \end{tabular*}
\end{table*}
\section{Optimal statistical significance}
\label{sec:oot}
The optimal observable technique (OOT)~\cite{Atwood:1991ka,Gunion:1996vv,Diehl:1993br,Davier:1992nw,Bhattacharya:2021ltd} is a powerful method to estimate model parameters or couplings from experimental data with maximum statistical sensitivity. 
By constructing observables that are analytically derived from the dependence of the differential cross-section on the parameters of interest, this approach 
ensures the most efficient use of the available kinematic information, by minimizing the covariance matrix. 
It is particularly useful in precision measurements, where small deviations from the SM predictions are expected to be estimated.

The differential cross section for any given collider process can be written in the form
    \begin{align}
    \mathcal{O}(\phi) &= \frac{d\sigma_{\rm}}{d\phi}\bigg|_{\tt observed} = \epsilon_{\tt S}\; \frac{d\sigma_{\rm}}{d\phi}\bigg|_{\tt S, theory} + \epsilon_{\tt B}\; \frac{d\sigma_{\rm}}{d\phi}\bigg|_{\tt B, theory} \nonumber\\&= \sum_{i}\; g_{i} f_{i}(\phi)\,,
    \end{align}
where $\phi$ is a phase space variable, $g_{i}$ are the functions containing the NP couplings and $f_{i}$ are the functions of phase space variable, $\phi$. The subscripts ${\tt S}$ and ${\tt B}$ refers to signal and background, respectively. $\epsilon_{\tt S}$ and $\epsilon_{\tt B}$ are the signal and background detection efficiencies, following all the cuts and 
final state signal identification. The optimal covariance matrix upon minimization ($\partial V_{ij}/\partial g_{j}=0$) is obtained as,
\begin{align}
    V_{ij} = \frac{M^{-1}_{ij}}{\mathcal{L}_{\rm int}} = \frac{1}{\mathcal{L}_{\rm int}}\int \frac{f_i(\phi)f_j(\phi)}{\mathcal{O}(\phi)}d\phi.
\end{align}
Correspondingly optimal $\chi^{2}$ is given by
\begin{align}
    \chi^{2} = \sum_{i,j}\; (g_{i}-g^{0}_{i})\;V^{-1}_{ij}\;(g_{j}-g^{0}_{j})\; \bigg|_{g=g^{0}}\,,
\end{align}
where $g^{0}$ correspond to the seed values of the coefficients $g$. The input seed values can come from different sources like available measurements, 
predictions from a different experiment, etc. In case of sensitivity prediction of an unobserved NP scenario, the seed value is usually determined by setting 
the seed values of NP couplings to zero.

For $e^+e^- \to Zh$ production at the ILC, we map the accessible phase space using the variable, $\cos{\theta}$, which is the emerging angle of the $Z$ boson. The differential observable can be written in terms of polarization degrees as
\begin{widetext}
    \begin{align}
    \mathcal{O}(\cos{\theta}) = \epsilon_{\tt S}\; \frac{d\sigma}{d\cos{\theta}} \bigg|_{\tt S, theory}
    &\;= \epsilon_{\tt S}\; \left\{ g^{L}_{0}\, (1-\overline{P_{e^+}})(1-P_{e^-}) + g^{R}_{0}\, (1+\overline{P_{e^+}})(1+P_{e^-}) \right\} f_{0} \nonumber\\
    &\;+\; \epsilon_{\tt S}\; \left\{ g^{L}_{1}\, (1-\overline{P_{e^+}})(1-P_{e^-}) + g^{R}_{1}\, (1+\overline{P_{e^+}})(1+P_{e^-}) \right\} f_{1} \nonumber\\
    &\;+\; \epsilon_{\tt S}\; \left\{ g^{L}_{2}\, (1-\overline{P_{e^+}})(1-P_{e^-}) + g^{R}_{2}\, (1+\overline{P_{e^+}})(1+P_{e^-}) \right\} f_{2}\;. \nonumber\\
    \label{eq:gf}
    \end{align}
\end{widetext}
The background contribution is small after the signal selection cuts are applied, as seen from Sec. \ref{sec:ml}, and this has been neglected to estimate 
optimal sensitivity. The linearly independent phase space functions $f_{i}$ are defined as follows:
\begin{equation}
    f_{0} = \beta\,, \hspace{1cm} f_{1} = \beta \cos{\theta}\,, \hspace{1cm} f_{2} = \beta \cos^{2}{\theta}\,, 
\end{equation}
where,
\begin{equation}
    \beta = \frac{\sqrt{\lambda(s, m^{2}_{Z}, m^{2}_{H})}}{32 \pi s^{2}}\,.
\end{equation}
Here, $\lambda$ is the Kallen function defined as $\lambda(x,y,z)= x^2 + y^2 + z^2 - 2xy - 2yz - 2zx$. With $\sqrt{s}=250$ GeV, $m_{Z} = 91$ GeV and $m_{h} = 125$ GeV, we obtain: $\beta \approx 31$ pb. With all the SM constants replaced by their respective values, the dimensionless $g_i$ as in Eq.~\ref{eq:gf}, turns out to be functions of the WCs 
only. We list the expressions for $g_i$ in  Appendix \ref{sec:oot-couplings}. Given that there is no hint of BSM, we choose the seed values of all the WCs to be zero to find an 
optimal range in which they can be extracted. 

\section{Results}
\label{sec:result}

In this section, we present projected sensitivities to the dimension-6 operators in Eq.~\eqref{eq:dim6}, derived using the optimal observable technique. Constraints on the corresponding Wilson coefficients (WCs) are obtained following the optimized signal selection strategy described earlier. We do the analysis for various beam polarization 
configurations, highlighting the improvement in sensitivity enabled by polarized $e^+e^-$ collisions. For benchmarking, we also compare our projections with existing 
limits from the LHC and provide estimates relevant to future collider runs.

Table~\ref{tab:lhcvsilc} summarizes the optimal 68\% CL limits (corresponding to $\chi^2 = 1$) on the WCs at the ILC with $\sqrt{s} = 250$ GeV, 
in comparison with current CMS constraints at 13 TeV with $\mathcal{L}_{\rm int} = 138~{\rm fb}^{-1}$. Two ILC scenarios are considered: (I) unpolarized beams with 
$\mathcal{L}_{\rm int} = 138~{\rm fb}^{-1}$, and (II) a polarized setup combining $(+30\%,-80\%)$ and $(-30\%,+80\%)$ beam configurations, each contributing $69~{\rm fb}^{-1}$, 
summing to the same total integrated luminosity.
\begin{table*}[htb!] 
    \centering
    \caption{68\% CL bounds on CP even (\textit{top half}) and CP odd (\textit{bottom half}) operator coefficients at the LHC vs. from $Zh$ production at the ILC 250 GeV. For polarized case, we choose $\mathfrak{L}_{\rm int} =$ 69 fb$^{-1}$ for polarization setups $(+30\%,-80\%)$ and $(-30\%,+80\%)$ each.}
    \label{tab:lhcvsilc}
     \renewcommand{\arraystretch}{1.5}
		\begin{tabular*}{1\textwidth}{@{\extracolsep{\fill}}cccc@{}}
    \hline \hline
    \multirow{2}*{WCs} & \multirow{2}*{LHC (13 TeV 138 fb$^{-1}$)} & \multicolumn{2}{c}{ILC (250 GeV 138 fb$^{-1}$)} \\ \cline{3-4}
    & & Unpolarized & Polarized $(\pm30\%,\mp80\%)$ \\
    \hline \hline
    $C_{H W}$ & $[-0.79, +0.51]$ & $[-0.11, +0.11]$ & $[-0.09, +0.08]$ \\
    $C_{H WB}$ & $[-1.62, +1.50]$ & $[-0.24, +0.22]$ & $[-0.16, +0.15]$ \\
    $C_{H B}$ & $[-0.23, +0.16]$ & $[-0.69, +0.41]$ & $[-0.14, +0.14]$ \\ \hline 
    $C_{H \widetilde{W}}$ & $[-0.76, +0.41]$ & $[-1.36, +1.36]$ & $[-1.16, +1.16]$ \\ 
    $C_{H \widetilde{W} B}$ & $[-1.57, +0.83]$ & $[-2.50, +2.50]$ & $[-1.98, +1.98]$ \\ 
    $C_{H \widetilde{B}}$ & $[-0.23, +0.12]$ & $[-2.16, +2.16]$ & $[-1.84, +1.84]$ \\
    \hline \hline
    \end{tabular*}
\end{table*}
The sensitivity to the CP-even operator \(C_{HW}\) improves by approximately a factor of 6 in the unpolarized ILC scenario compared to the current LHC constraints, with beam polarization providing an additional enhancement to about a factor of 7. For \(C_{HWB}\), the improvement is even more pronounced, reaching nearly a factor of 7 in the unpolarized case and exceeding a factor of 10 when polarized beams are employed. In the case of \(C_{HB}\), the unpolarized ILC limit is slightly weaker than the LHC bound; however, the introduction of beam polarization yields an improvement of roughly 1.5 times. These results highlight the critical role of beam polarization in maximizing the precision reach of future lepton colliders.

For the CP-odd operators, the current LHC limits generally demonstrate stronger sensitivity compared to the ILC projections obtained via the optimal observable technique. This reduced sensitivity at the ILC arises because the optimal observable method probes CP-odd effects predominantly at the quadratic level in the differential cross section, which inherently suppresses the constraints. As reflected in Table~\ref{tab:lhcvsilc}, the allowed parameter ranges for these operators remain significantly wider at the ILC, even with beam polarization. To enhance sensitivity to CP-odd operators, alternative strategies---such as leveraging spin-polarization asymmetries or dedicated CP-odd observables---could provide more powerful and direct probes, thus offering complementary avenues to tighten these bounds at future lepton colliders. A detailed study exploiting these observables to constrain CP-odd operators will be presented in future work.

\begin{figure*}[htb!]
    \centering
    \includegraphics[width=0.32\linewidth]{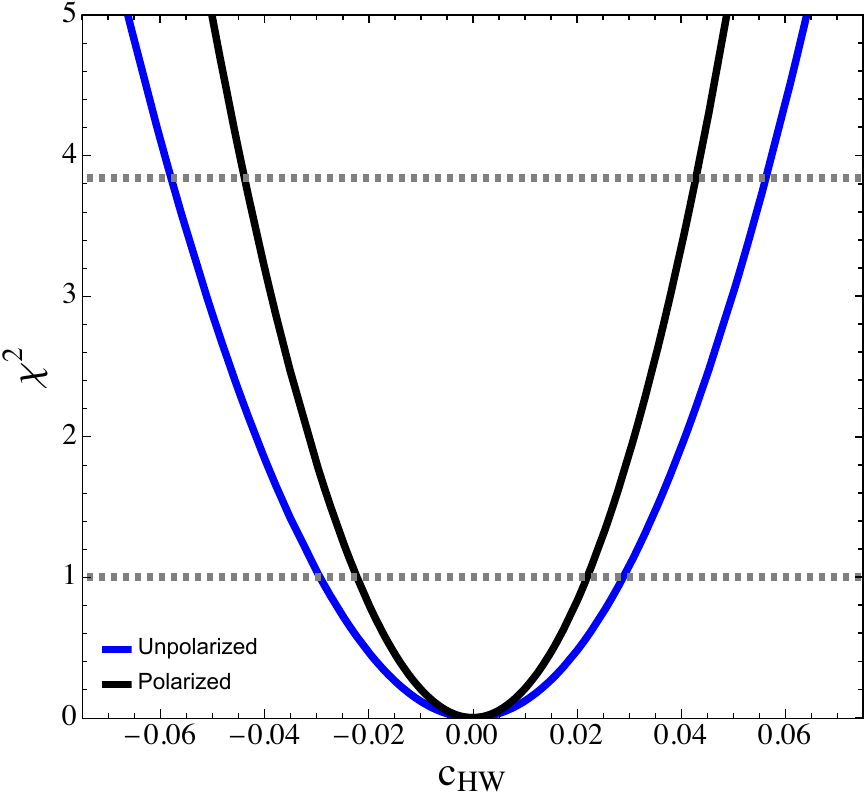}
    \includegraphics[width=0.32\linewidth]{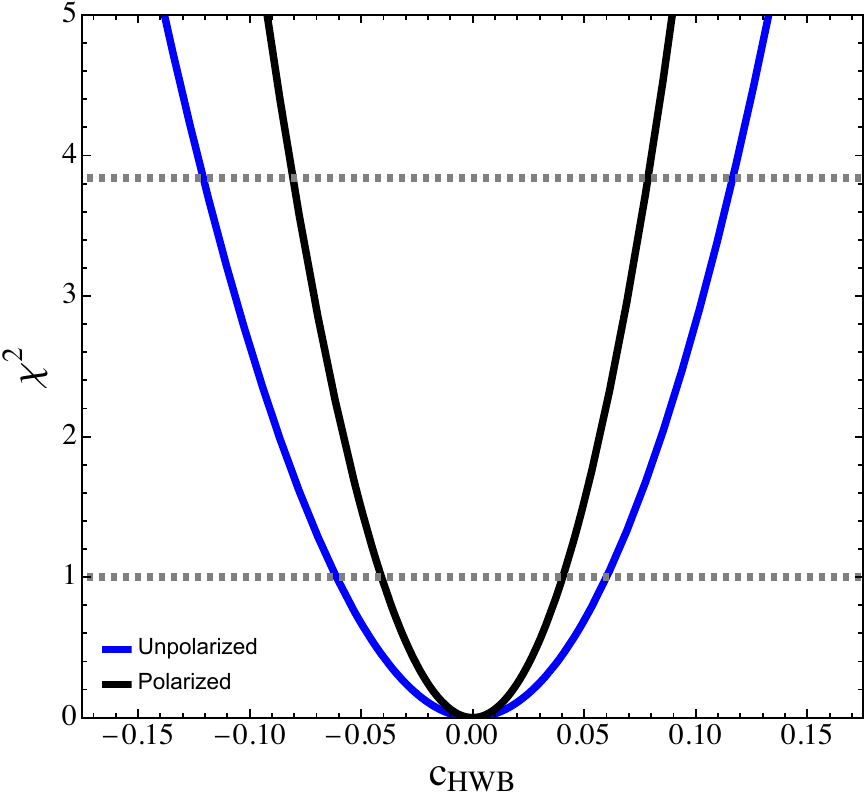}
    \includegraphics[width=0.32\linewidth]{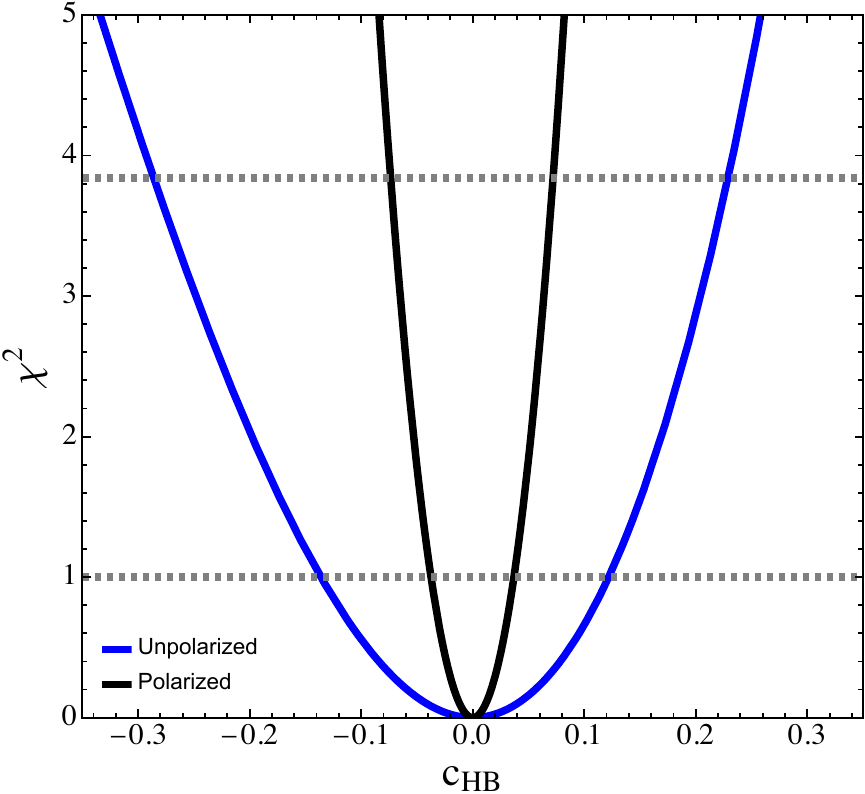}
    \includegraphics[width=0.32\linewidth]{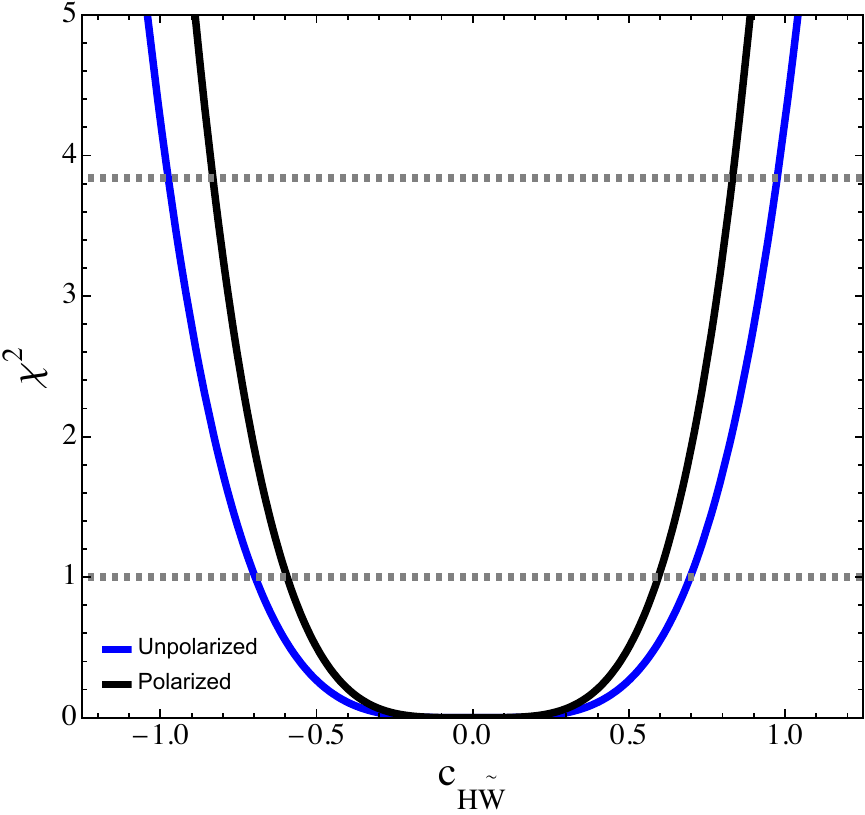}
    \includegraphics[width=0.32\linewidth]{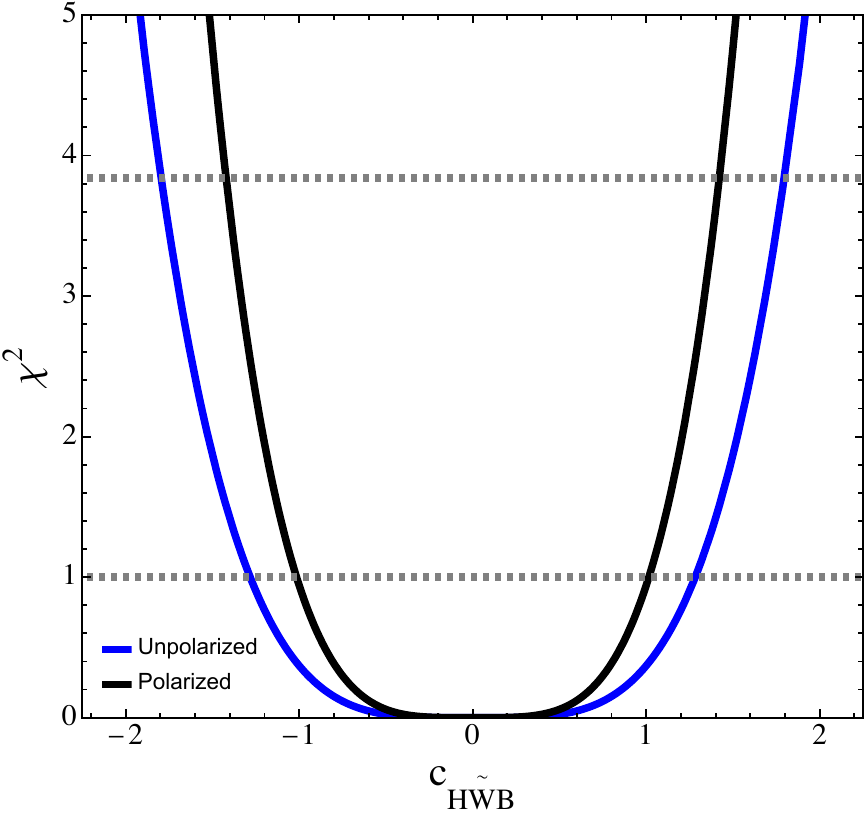}
    \includegraphics[width=0.32\linewidth]{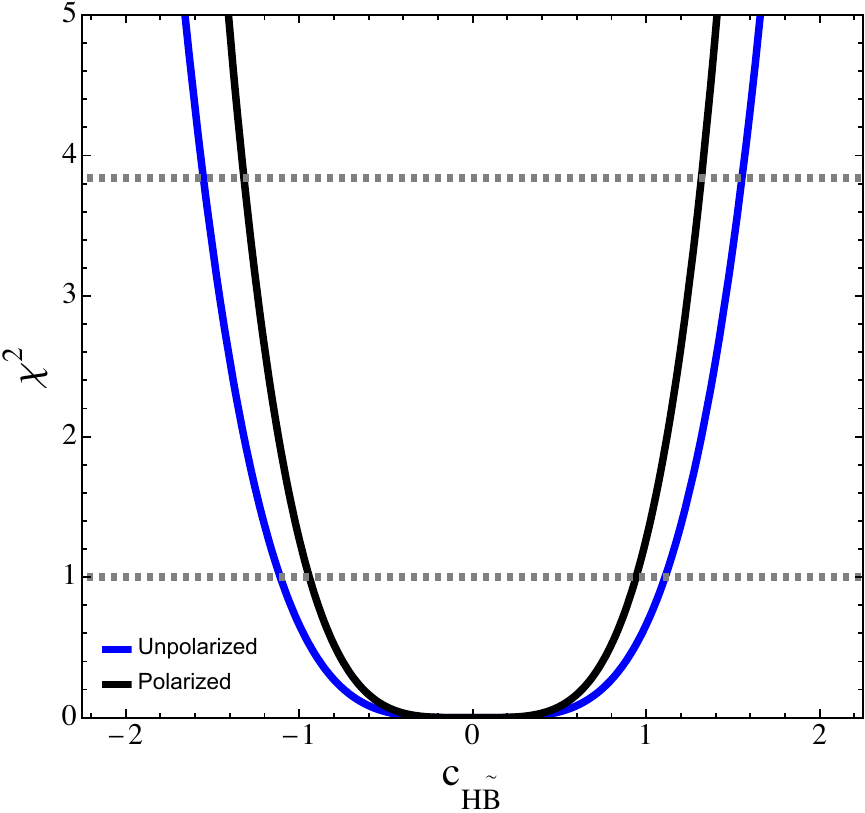}
    \caption{One parameter optimal sensitivity plots from $Zh$ production at the ILC 250 GeV for unpolarized and polarized setups. For the unpolarized case, $\mathcal{L}_{\rm int}=$ 2000 fb$^{-1}$, and for the polarized case each polarization setup i.e. $(+30\%,-80\%)$ and $(-30\%,+80\%)$ corresponds to $\mathcal{L}_{\rm int}=$ 1000 fb$^{-1}$, hence giving a combined luminosity, $\mathcal{L}_{\rm int}=$ 2000 fb$^{-1}$.}
    \label{fig:oot1d}
\end{figure*}

\begin{table*}[htb!]
    \centering
    \caption{Projected sensitivities of CP even (\textit{top half}) and CP odd (\textit{bottom half}) SMEFT operator coefficients from $Zh$ production at the ILC 250 GeV. For the polarized setup, we combine polarization setups: $(+30\%,-80\%)$ and $(-30\%,+80\%)$, each with luminosities, $\mathcal{L}_{\rm int}=$ 1000 fb$^{-1}$, hence the combined setup of $\mathcal{L}_{\rm int}=$ 2000 fb$^{-1}$.}
    \label{tab:oots1}
    \renewcommand{\arraystretch}{1.25}
    \begin{tabular*}{1\textwidth}{@{\extracolsep{\fill}}ccccccc@{}}
        \hline \hline
        \multirow{3}*{WCs} & \multirow{3}*{C.L.} & \multicolumn{5}{c}{Projected sensitivities} \\ \cline{3-7}
        & & \multicolumn{2}{c}{Unpolarized} & $(+30\%,-80\%)$ & $(-30\%,+80\%)$ & Polarized \\ \cline{3-4}
        & & $\mathfrak{L}_{\rm int}=$ 1000 fb$^{-1}$ & $\mathfrak{L}_{\rm int}=$ 2000 fb$^{-1}$ & $\mathfrak{L}_{\rm int}=$ 1000 fb$^{-1}$ & $\mathfrak{L}_{\rm int}=$ 1000 fb$^{-1}$ & $\mathfrak{L}_{\rm int}=$ 2000 fb$^{-1}$ \\
        \hline \hline
        \multirow{2}*{$C_{H W}$} & 68\% & $[-0.04, +0.04]$ & $[-0.03, +0.03]$ & $[-0.02, +0.02]$ & $[-0.17, +0.16]$ & $[-0.02, +0.02]$ \\
        & 95\% & $[-0.08, +0.08]$ & $[-0.06, +0.06]$ & $[-0.04, +0.04]$ & $[-0.34, +0.31]$ & $[-0.04, +0.04]$ \\ \hline
        \multirow{2}*{$C_{H W B}$} & 68\% & $[-0.09, +0.08]$ & $[-0.06, +0.06]$ & $[-0.33, +0.31]$ & $[-0.04, +0.04]$ & $[-0.04, +0.04]$ \\
        & 95\% & $[-0.17, +0.16]$ & $[-0.12, +0.12]$ & $[-0.66, +0.60]$ & $[-0.08, +0.08]$ & $[-0.08, +0.08]$ \\ \hline
        \multirow{2}*{$C_{H B}$} & 68\% & $[-0.20, +0.17]$ & $[-0.14, +0.12]$ & $[-0.10, +0.10]$ & $[-0.04, +0.04]$ & $[-0.04, +0.04]$ \\
        & 95\% & $[-0.44, +0.31]$ & $[-0.28, +0.23]$ & $[-0.19, +0.20]$ & $[-0.08, +0.08]$ & $[-0.07, +0.07]$ \\ \hline \hline
        \multirow{2}*{$C_{H \widetilde{W}}$} & 68\% & $[-0.83, +0.83]$ & $[-0.70, +0.70]$ & $[-0.60, +0.60]$ & $[-2.18, +2.18]$ & $[-0.59, +0.59]$ \\
        & 95\% & $[-1.16, +1.16]$ & $[-0.98, +0.98]$ & $[-0.83, +0.83]$ & $[-3.05, +3.05]$ & $[-0.83, +0.83]$ \\ \hline
        \multirow{2}*{$C_{H \widetilde{W} B}$} & 68\% & $[-1.52, +1.52]$ & $[-1.28, +1.28]$ & $[-4.28, +4.28]$ & $[-1.02, +1.02]$ & $[-1.02, +1.02]$ \\
        & 95\% & $[-2.14, +2.14]$ & $[-1.80, +1.80]$ & $[-5.99, +5.99]$ & $[-1.42, +1.42]$ & $[-1.42, +1.42]$ \\ \hline
        \multirow{2}*{$C_{H \widetilde{B}}$} & 68\% & $[-1.32, +1.32]$ & $[-1.11, +1.11]$ & $[-2.04, +2.04]$ & $[-0.95, +0.95]$ & $[-0.94, +0.94]$ \\
        & 95\% & $[-1.85, +1.85]$ & $[-1.55, +1.55]$ & $[-2.86, +2.86]$ & $[-1.33, +1.33]$ & $[-1.32, +1.32]$ \\ \hline \hline
    \end{tabular*}
\end{table*}

We further study the projected sensitivities of the relevant SMEFT operator coefficients at the ILC with an integrated luminosity of 1000 fb$^{-1}$ per polarization configuration, and a combined dataset yielding 2000 fb$^{-1}$ luminosity. The one-parameter sensitivity projections are shown in Fig.~\ref{fig:oot1d} for graphical representation. The gray dashed lines indicate the 68\% confidence level (CL) i.e., $\chi^2 = 1$ and 95\% CL ($\chi^2 = 3.84$) thresholds. The corresponding numerical limits are listed in Tab. \ref{tab:oots1}. For the CP-even operators, the results demonstrate notable variations in sensitivity across the different configurations. The $C_{H W}$ operator benefits significantly from beam polarization; the $(+30\%,-80\%)$ setup yields the most stringent bounds at both 68\% and 95\% confidence levels, improving roughly by a factor of two compared to the unpolarized case. The $(-30\%,+80\%)$ configuration shows weaker sensitivity for this operator, illustrating the asymmetric impact of polarization states. 

In the case of $C_{HWB}$, the situation is reversed with the $(-30\%,+80\%)$ polarization providing the strongest limits, surpassing both the unpolarized and the $(+30\%,-80\%)$ runs by a considerable margin. This complementarity between polarized states underscores the importance of running both configurations for a comprehensive probe. The combined analysis again improves the bounds further, achieving nearly a factor of two improvement over single configurations. For $C_{H B}$, the $(-30\%,+80\%)$ polarization scenario delivers the best sensitivity, with bounds tighter by roughly a factor of four compared to the unpolarized run. 

Turning to the CP-odd operators, the sensitivities exhibit a more varied pattern. For $C_{H \widetilde{W}}$, the $(+30\%,-80\%)$ polarization setup outperforms the others, improving constraints by approximately $30\%$ relative to unpolarized collisions. The flipped polarization configuration is less sensitive.

For $C_{H\widetilde{W}B}$ and $C_{H\widetilde{B}}$, the $(-30\%,+80\%)$ run provides stronger bounds than either unpolarized or flipped configuration individually. However, combining all data still leads to a modest but meaningful improvement, demonstrating the statistical benefit of aggregating diverse polarization states. 

Overall, these projections emphasize the critical role of beam polarization in enhancing sensitivity to dimension-6 operators at the ILC, with different operators responding distinctly to polarization choices. The combined dataset invariably delivers the most stringent bounds, leveraging the complementarity and statistical gains across all configurations.

We further compare our results with Ref.~\cite{Ogawa:2017bmg}, which analyzed $e^+e^- \to Zh$ at $\sqrt{s}=250$~GeV under identical beam polarizations and an integrated luminosity of $2000~\text{fb}^{-1}$. That study parameterized the anomalous $hZZ$ vertex in terms of three real couplings, $a_Z$, $b_Z$, and $\widetilde{b}_Z$, extracted from angular and cross~section observables. In the SMEFT framework, our dimension-6 operators contribute only to $b_Z$ and $\widetilde{b}_Z$. Inverting the couplings, we obtain $b_Z \in [-0.0079,\,+0.0079]$ and $\bar{b}_Z \in [-1.517,\,+1.517]$. The bound on the CP-even parameter $b_Z$ agrees well with Ref.~\cite{Ogawa:2017bmg}, while the much weaker limit on the CP-odd parameter $\bar{b}_Z$, about two orders of magnitude larger, reflects the limited CP sensitivity of total cross-section observables and underscores the need for CP-odd and spin-correlation analyses, as discussed above.

\begin{figure*}[htb!]
    \centering
    \includegraphics[width=0.32\linewidth]{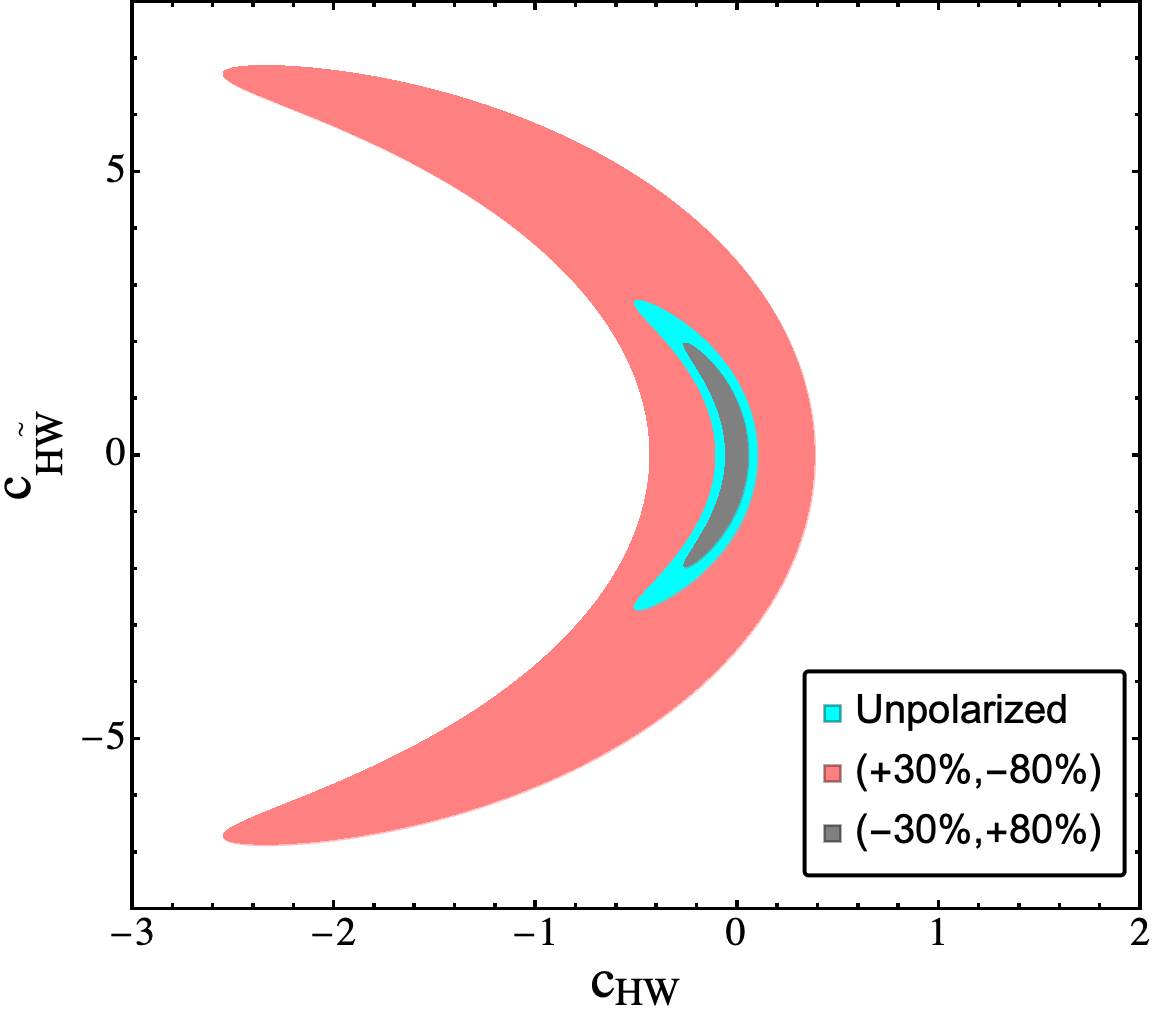}
    \includegraphics[width=0.32\linewidth]{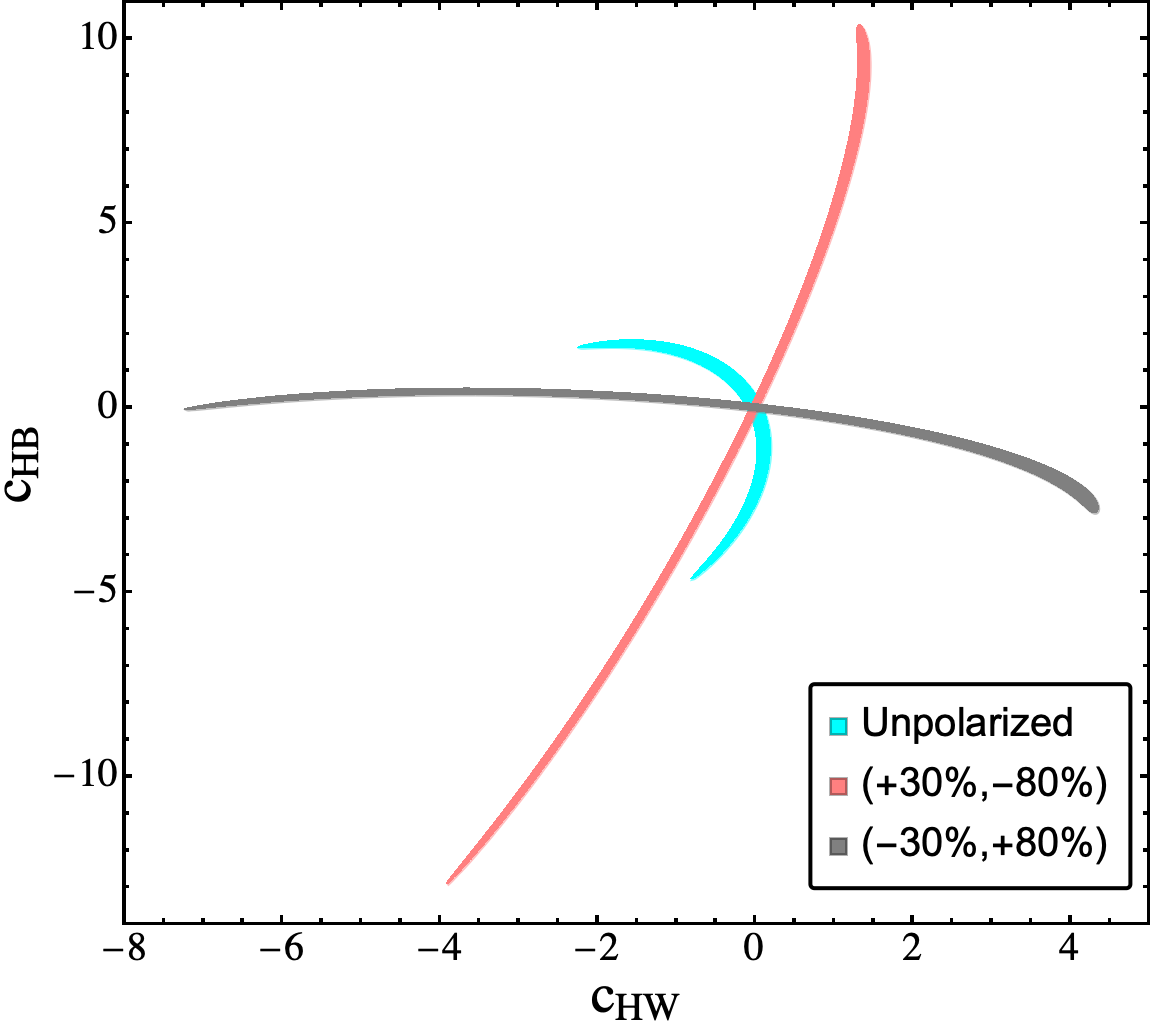}
    \includegraphics[width=0.32\linewidth]{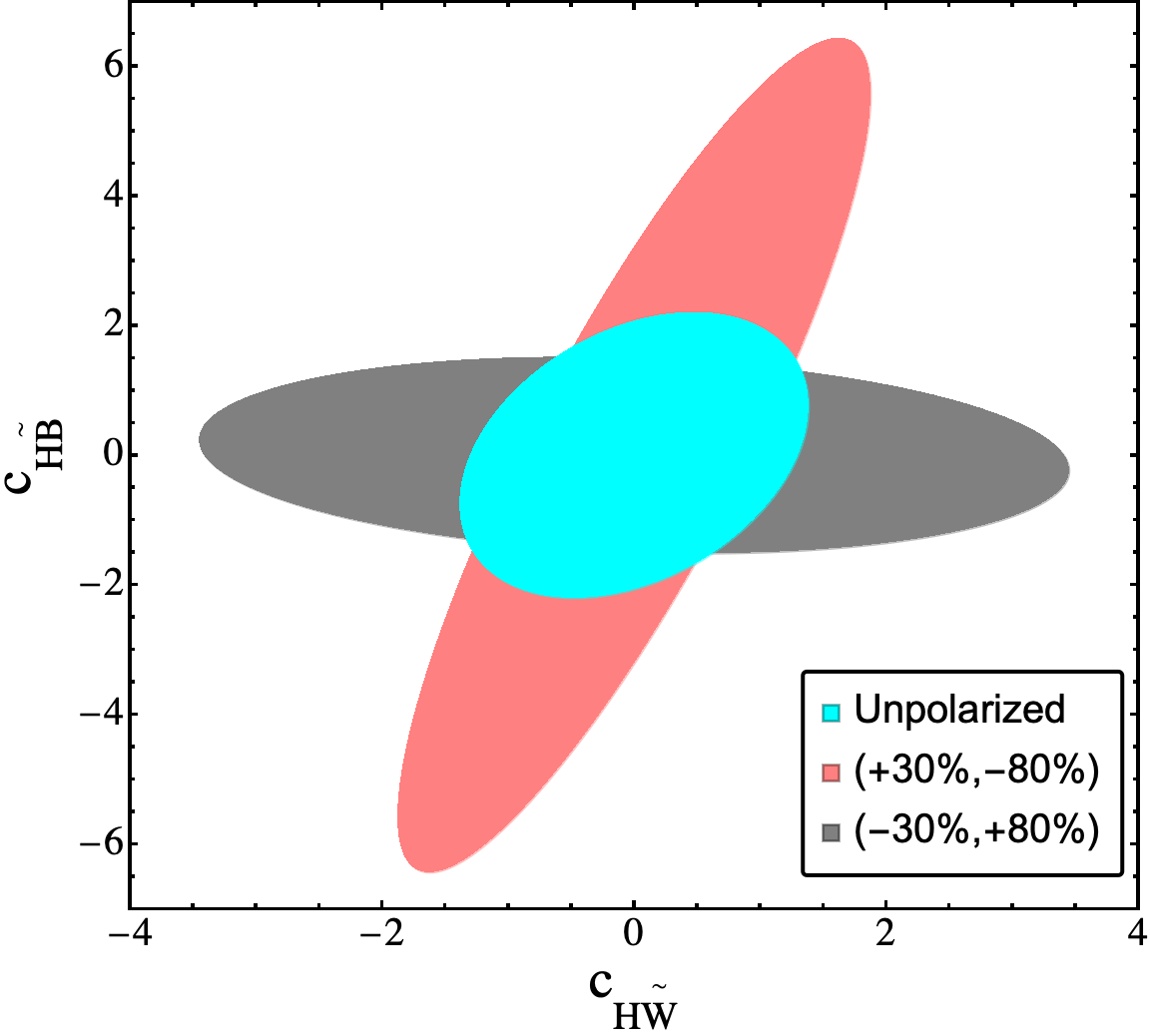}
    \caption{Two parameter 95\% CL optimal sensitivity plots from $Zh$ production at the ILC 250 GeV for unpolarized and different polarization setups. For each setup, $\mathcal{L}_{\rm int}=$ $1000$ fb$^{-1}$.}
    \label{fig:2dpol}
\end{figure*}

In addition, the two parameter sensitivity contours are illustrated in Fig. \ref{fig:2dpol}, providing a complementary view of the correlations and degeneracies between operator coefficients that are not captured in the one parameter analysis. Additional sensitivity contours are presented in Appendix \ref{sec:2doot}.

\section{Conclusion}
\label{sec:conclude}

The associated production of a single Higgs boson with a $Z$ boson—commonly referred to as the Higgs-strahlung process—offers a clean and model-independent channel to probe Higgs-gauge couplings with high precision. In this study, we focus on the final state consisting of two oppositely charged, same-flavor leptons and two $b$-tagged jets in 
$e^+e^-$ collisions at a center-of-mass energy of $\sqrt{s} = 250$~GeV, incorporating scenarios with polarized initial beams. The deviations from the Standard Model (SM) expectations are parameterized by the SMEFT framework, allowing us to constrain the coefficients of higher-dimensional operators that modify the $hZZ$ and $hZ\gamma$ interactions.

Event selection is performed via the recoil mass technique, targeting the Higgs boson through the invariant mass spectrum of the $Z$-tagged dilepton system. 
The statistical extraction of the sensitivity to the Wilson coefficients (WCs) is carried out using the Optimal Observable Technique (OOT), which maximizes information 
from kinematic distributions by exploiting interference patterns between SM and SMEFT amplitudes.

Our analysis demonstrates that beam polarization plays a crucial role in enhancing sensitivity to various operators. When comparing our results with current bounds from the 
CMS experiment at $\sqrt{s} = 13$~TeV and $\mathcal{L}_{\rm int} = 138$~fb$^{-1}$, we find that the ILC setup, with polarized beams, yields significantly tighter 
constraints—improving the bounds on CP-even operators by factors ranging from $1.5$ to $10$. This improvement stems from the linear interference term between the 
SM and the CP-even SMEFT contributions, which scales as $1/\Lambda^2$ and benefits strongly from the clean environment and polarization of the $e^+e^-$ collider.

On the other hand, the sensitivity to CP-odd operators is comparatively weaker than that of CMS. This is attributed to the fact that CP-odd operators contribute dominantly 
at quadratic order in $1/\Lambda^2$, resulting in suppressed effects at the amplitude level. Consequently, our observable—being inclusive in nature—lacks sensitivity to 
CP-odd interference effects. This highlights the need for a more targeted analysis involving CP-sensitive observables such as angular asymmetries, triple-product correlations, 
or differential distributions optimized for CP-violation.

Finally, we present projections at higher integrated luminosity of $\mathcal{L}_{\rm int} = 2000$~fb$^{-1}$ for two different ILC running scenarios: an unpolarized beam and a combination of
two complementary polarized beam configurations $(+30\%,-80\%)$ and $(-30\%,+80\%)$, with integrated luminosity of $\mathcal{L}_{\rm int} = 2000$~fb$^{-1}$ each. We observe that different polarization choices enhance sensitivity to different 
subsets of operators due to their distinct helicity structures. The combination of the two polarization configurations, with $\mathcal{L}_{\rm int} = 2000$~fb$^{-1}$, yields the strongest bounds across the operator basis effectively exploiting the polarization-enhanced interference terms. These results reinforce the power of polarized $e^+e^-$ collisions at the ILC in probing new physics in the Higgs sector with high precision.

\begin{acknowledgments}
A Subba and S Bhattacharya acknowledges the ANRF grant CRG/2023/000580, under which A Subba is supported.
\end{acknowledgments}

\appendix
\section{ML-based insights}
\label{sec:bdt}
\begin{figure}[htb!]
    \centering
    \includegraphics[width=0.95\linewidth]{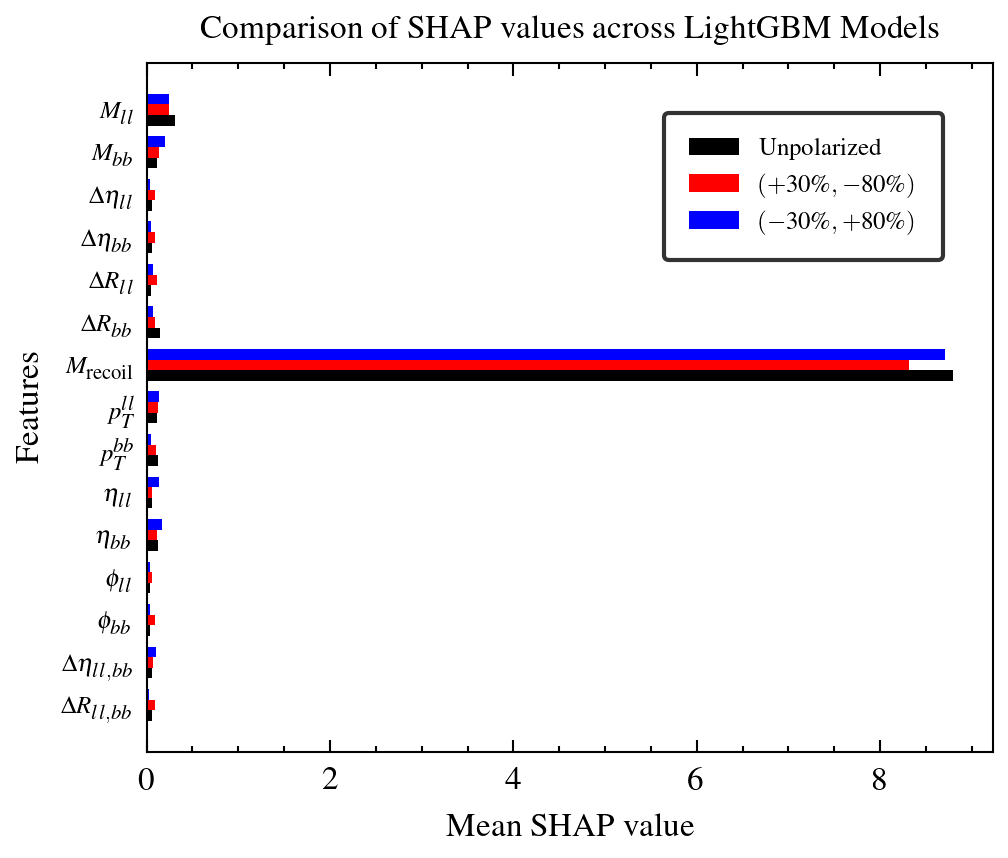}
    \caption{SHAP values for BDT models trained for signal-background segregation.}
    \label{fig:shap}
\end{figure}
We train three BDT models using the \texttt{LightGBM} framework: one for the unpolarized case and two for the polarized beam configurations $(P_{e^+}, P_{e^-}) = (+30\%, -80\%)$ and $(-30\%, +80\%)$. For each model, the dataset is split into equal halves for training and testing. The kinematic variables (features) used for training are as follows: Invariant mass of the dilepton system ($M_{ll}$) and the $b\bar{b}$ system ($M_{bb}$), recoil mass of the dilepton system ($m_{\rm Recoil}$), angular separation between the leptons: $\Delta \eta_{ll}$, $\Delta R_{ll}$, angular separation between the $b$-jets: $\Delta \eta_{bb}$, $\Delta R_{bb}$, kinematic variables of the dilepton system: transverse momentum ($p_{T}^{ll}$), pseudorapidity ($\eta_{ll}$), and azimuthal angle ($\phi_{ll}$), kinematic variables of the $b\bar{b}$ system: $p_{T}^{bb}$, $\eta_{bb}$, and $\phi_{bb}$, angular separation between the dilepton and $b\bar{b}$ systems: $\Delta \eta_{ll,bb}$, $\Delta R_{ll,bb}$. The recoil mass of the dilepton system is defined as:
\begin{equation}
    m_{\rm Recoil} = \sqrt{s - 2\sqrt{s}E_{ll} + M_{ll}^2} \;,
\end{equation}
where $E_{ll}$ is the total energy of the two leptons. The definitions of the other variables follow standard usage in collider physics literature. The polar angle of the final state, reconstructed from the dilepton system, is defined as:
\begin{equation}
    \cos{\theta}_{\rm Recoil} = \frac{p_{z}^{ll}}{|\vec{p}_{ll}|}
\end{equation}
where  $p_{z}^{ll}$ is the longitudinal momentum of the dilepton system and $|\vec{p}_{ll}|$ is its total three-momentum magnitude. All input features are normalized using the \texttt{StandardScaler} module. The BDT models are trained with equal class weights for signal and background events, while testing is performed using weights proportional to the true event yields. The SHAP values for the trained models are shown in Fig. \ref{fig:shap}. Among all features, the recoil mass $m_{\rm Recoil}$ emerges as the most powerful discriminator. Therefore, we choose to replace the BDT model with a simple optimized selection cut on $m_{\rm Recoil}$, achieving similar classification performance with greater interpretability.

\section{Couplings in $Zh$ differential cross-section as Optimal Observable}
\label{sec:oot-couplings}

As stated before, the differential cross-section for $e^+e^- \to Zh$ serves as the observable to estimate the NP couplings (WCs) optimally in ILC. In the limit, where background contamination are assumed negligible, the observable can be expressed as, 
\begin{equation}
    \begin{aligned}
    \mathcal{O}(\phi) &= \frac{d\sigma_{\rm}}{d\phi}\bigg|_{\tt observed} = \epsilon_{\tt S}\; \frac{d\sigma_{\rm}}{d\phi}\bigg|_{\tt S, theory} \\&= \sum_{i}\; g_{i} f_{i}(\phi)\,,
    \end{aligned}
\end{equation} 
Upon a semi-numeric evaluation, with $\sqrt{s}=250$ GeV, $m_{Z} = 91$ GeV and $m_{h} = 125$ GeV, the couplings attached to the phase space variables as in 
Eq.~\ref{eq:gf}, are furnished in Eq.~\eqref{eq:gi1}.
\begin{widetext}
    \begin{align}
        \label{eq:gi1}
        g^{L}_{0} &=\;\;\;\; 2.0\times 10^{-3} + 1.8 \times 10^{-3}\; C_{H W} + 7.7 \times 10^{-5}\; C_{H WB} + 4.4 \times 10^{-4}\; (C_{H W})^2 
        - 4.7 \times 10^{-4}\; (C_{H B})^2 \nonumber\\&+ 7.7 \times 10^{-7}\; (C_{H WB})^2 - 2.3 \times 10^{-4}\; C_{H W}\,C_{H B}
        - 9.4 \times 10^{-6}\; C_{H B}\,C_{H WB} + 8.4 \times 10^{-5}\; (C_{H \widetilde{W}})^2 \nonumber\\&+ 5.4 \times 10^{-6}\; (C_{H \widetilde{B}})^2 
        + 1.5 \times 10^{-7}\; (C_{H \widetilde{W}B})^2 - 4.2 \times 10^{-5}\; C_{H \widetilde{W}}\,C_{H \widetilde{B}} + 7.0 \times 10^{-7}\; C_{H \widetilde{W}}\,C_{H \widetilde{W}B}\nonumber \\
        &- 1.8 \times 10^{-6}\; C_{H \widetilde{B}}\,C_{H \widetilde{W}B}\,,\nonumber\\
         g^{L}_{0} &=\;\;\;\; 2.0\times 10^{-3} + 1.8 \times 10^{-3}\; C_{H W} + 7.7 \times 10^{-5}\; C_{H WB} + 4.4 \times 10^{-4}\; (C_{H W})^2 
        - 4.7 \times 10^{-4}\; (C_{H B})^2 \nonumber\\&+ 7.7 \times 10^{-7}\; (C_{H WB})^2 - 2.3 \times 10^{-4}\; C_{H W}\,C_{H B}
        - 9.4 \times 10^{-6}\; C_{H B}\,C_{H WB} + 8.4 \times 10^{-5}\; (C_{H \widetilde{W}})^2 \nonumber\\&+ 5.4 \times 10^{-6}\; (C_{H \widetilde{B}})^2 
        + 1.5 \times 10^{-7}\; (C_{H \widetilde{W}B})^2 - 4.2 \times 10^{-5}\; C_{H \widetilde{W}}\,C_{H \widetilde{B}} + 7.0 \times 10^{-7}\; C_{H \widetilde{W}}\,C_{H \widetilde{W}B} \nonumber\\
        &- 1.8 \times 10^{-6}\; C_{H \widetilde{B}}\,C_{H \widetilde{W}B}\,,\nonumber\\
        g^{L}_{1} &=\;\;\;\;3.2 \times 10^{-19}\; C_{H W} - 3.2 \times 10^{-19}\; C_{H B} - 5.4 \times 10^{-20}\; C_{H WB} 
        - 5.4 \times 10^{-20}\; (C_{H W})^2 - 2.7 \times 10^{-20}\; (C_{H B})^2 \nonumber\\&- 1.3 \times 10^{-19}\; C_{H W}\,C_{H B}
        - 4.7 \times 10^{-20}\; C_{H W}\,C_{H WB} + 2.4 \times 10^{-20}\; (C_{H \widetilde{W}})^2 + 2.4 \times 10^{-20}\; (C_{H \widetilde{B}})^2\nonumber \\
        &+ 2.5 \times 10^{-21}\; (C_{H \widetilde{W}B})^2 - 4.7 \times 10^{-20}\; C_{H \widetilde{W}}\,C_{H \widetilde{B}} - 1.4 \times 10^{-20}\; C_{H \widetilde{W}}\,C_{\Phi \widetilde{W}B} + 1.4 \times 10^{-20}\; C_{\Phi \widetilde{B}}\,C_{\Phi \widetilde{W}B}\,,\nonumber\\
        g^{L}_{2} &=- 3.7 \times 10^{-4} - 5.4 \times 10^{-20}\; C_{\Phi W} + 5.4 \times 10^{-20}\; C_{\Phi B} + 1.4 \times 10^{-20}\; C_{\Phi WB} 
        + 8.4 \times 10^{-5}\; (C_{\Phi W})^2\nonumber\\& - 5.4 \times 10^{-6}\; (C_{\Phi B})^2 - 1.5 \times 10^{-7}\; (C_{\Phi WB})^2 
        - 4.3 \times 10^{-5}\; C_{\Phi W}\,C_{\Phi B} - 7.0 \times 10^{-6}\; C_{\Phi W}\,C_{\Phi WB} \nonumber\\&- 1.8 \times 10^{-6}\; C_{\Phi B}\,C_{\Phi WB} + 8.4 \times 10^{-5}\; (C_{\Phi \widetilde{W}})^2 + 5.4 \times 10^{-6}\; (C_{\Phi \widetilde{B}})^2 + 1.5 \times 10^{-7}\; (C_{\Phi \widetilde{W}B})^2]\nonumber \\
        &- 4.3 \times 10^{-5}\; C_{\Phi \widetilde{W}}\,C_{\Phi \widetilde{B}} - 7.0 \times 10^{-6}\; C_{\Phi \widetilde{W}}\,C_{\Phi \widetilde{W}B} + 1.8 \times 10^{-6}\; C_{\Phi \widetilde{B}}\,C_{\Phi \widetilde{W}B}\,,\nonumber\\
        g^{R}_{0} &= 1.4\times 10^{-3} + 1.0 \times 10^{-4}\; C_{H W} + 9.1 \times 10^{-4}\; C_{H B} + 8.6 \times 10^{-4}\; C_{H WB} 
+ 1.9 \times 10^{-6}\; (C_{H W})^2\nonumber \\&+ 1.5 \times 10^{-4}\; (C_{H B})^2 + 1.3 \times 10^{-4}\; (C_{H WB})^2 
+ 3.6 \times 10^{-7}\; (C_{H \widetilde{W}})^2 + 2.8 \times 10^{-5}\; (C_{H \widetilde{B}})^2 \nonumber\\&+ 2.5 \times 10^{-5}\; (C_{H \widetilde{W}B})^2 + 3.3 \times 10^{-5}\; C_{H W}\,C_{H B} + 3.2 \times 10^{-5}\; C_{H W}\,C_{H WB} \nonumber\\&+ 2.8 \times 10^{-4}\; C_{H B}\,C_{H WB} + 6.3 \times 10^{-6}\; C_{H \widetilde{W}}\,C_{H \widetilde{B}} + 6.0 \times 10^{-6}\; C_{H \widetilde{W}}\,C_{H \widetilde{W}B} + 5.2 \times 10^{-5}\; C_{H \widetilde{B}}\,C_{H \widetilde{W}B}\,,\nonumber\\
g^{R}_{1} &=- 1.1 \times 10^{-19}\; C_{H W} + 1.1 \times 10^{-19}\; C_{H B} + 1.1 \times 10^{-19}\; C_{H WB} - 5.4 \times 10^{-20}\; (C_{H W})^2 + 6.1 \times 10^{-20}\; (C_{H B})^2\nonumber \\&+ 2.0 \times 10^{-20}\; (C_{H WB})^2 + 2.4 \times 10^{-20}\; (C_{H \widetilde{W}})^2 + 1.0 \times 10^{-20}\; (C_{H \widetilde{B}})^2 + 2.5 \times 10^{-21}\; (C_{H \widetilde{W}B})^2\nonumber \\
&+ 1.4 \times 10^{-19}\; C_{H W}\,C_{H B} + 3.4 \times 10^{-20}\; C_{H W}\,C_{H WB} - 6.8 \times 10^{-21}\; C_{H B}\,C_{H WB} - 4.7 \times 10^{-20}\; C_{H \widetilde{W}}\,C_{H \widetilde{B}}\nonumber \\&- 1.4 \times 10^{-20}\; C_{H \widetilde{W}}\,C_{H \widetilde{W}B} + 2.7 \times 10^{-20}\; C_{H \widetilde{B}}\,C_{H \widetilde{W}B}\,,\nonumber\\
g^{R}_{2} &=- 2.8 \times 10^{-4} - 2.7 \times 10^{-20}\; C_{H W} + 2.7 \times 10^{-20}\; C_{H B} - 1.4 \times 10^{-20}\; C_{H WB} + 3.6 \times 10^{-7}\; (C_{H W})^2 \nonumber\\&+ 2.8 \times 10^{-5}\; (C_{H B})^2 + 2.5 \times 10^{-5}\; (C_{H WB})^2 + 3.6 \times 10^{-7}\; (C_{H \widetilde{W}})^2 + 2.8 \times 10^{-5}\; (C_{H \widetilde{B}})^2 \nonumber\\&+ 2.5 \times 10^{-5}\; (C_{H \widetilde{W}B})^2 + 6.3 \times 10^{-6}\; C_{H W}\,C_{H B} + 6.0 \times 10^{-6}\; C_{H W}\,C_{H WB} + 5.2 \times 10^{-5}\; C_{H B}\,C_{H WB} \nonumber\\
&+ 6.3 \times 10^{-6}\; C_{H \widetilde{W}}\,C_{H \widetilde{B}} + 6.0 \times 10^{-6}\; C_{H \widetilde{W}}\,C_{H \widetilde{W}B} + 5.2 \times 10^{-5}\; C_{H \widetilde{B}}\,C_{H \widetilde{W}B}\,.
        \end{align}
\end{widetext}

Note that they are quadratic in $C_i$ with cross-terms of different Wilson coefficients coming from different operators. The optimal $\chi^2$ analysis is done based on the above parametrization.
\section{Comparison with FCC-ee}
\label{sec:fcc-ee}
The electron-positron collision stage of the Future Circular Collider (FCC-ee)~\cite{FCC:2018evy} is expected to run at a CM energy of $\sqrt{s} = 240$ GeV with a projected integrated luminosity of $\mathcal{L}_{\rm int} = 10500$ fb$^{-1}$. However, the setup is expected to collide unpolarized beams only. In this section, we draw a comparison between the 95\% C.L. limits on the SMEFT operators for ILC 250 GeV with $\mathcal{L}_{\rm int} = 2000$ fb$^{-1}$ for the unpolarized and polarized setup discussed in Tab.~\ref{tab:oots1}, with the limits from the FCC-ee run, tabulated in Tab~\ref{tab:fccvsilc}. The higher luminosity of the FCC-ee results in more sensitive bounds compared to the ILC runs.

\begin{table*}[htb!] 
    \centering
    \caption{{\color{blue}95\% CL bounds on CP even (\textit{top half}) and CP odd (\textit{bottom half}) operator coefficients from $Zh$ production at the FCC-ee 240 GeV vs. ILC 250 GeV. For polarized case, we choose $\mathfrak{L}_{\rm int} =$ 1000 fb$^{-1}$ for polarization setups $(+30\%,-80\%)$ and $(-30\%,+80\%)$ each.}}
    \label{tab:fccvsilc}
     \renewcommand{\arraystretch}{1.5}
		\begin{tabular*}{1\textwidth}{@{\extracolsep{\fill}}cccc@{}}
    \hline \hline
    \multirow{2}*{WCs} & \multirow{2}*{FCC-ee (240 GeV 10500 fb$^{-1}$)} & \multicolumn{2}{c}{ILC (250 GeV 2000 fb$^{-1}$)} \\ \cline{3-4}
    & & Unpolarized & Polarized $(\pm30\%,\mp80\%)$ \\
    \hline \hline
    $C_{H W}$ & $[-0.02, +0.02]$ & $[-0.06, +0.06]$ & $[-0.04, +0.04]$ \\
    $C_{H WB}$ & $[-0.03, +0.03]$ & $[-0.12, +0.12]$ & $[-0.08, +0.08]$ \\
    $C_{H B}$ & $[-0.06, +0.06]$ & $[-0.28, +0.23]$ & $[-0.07, +0.07]$ \\ \hline 
    $C_{H \widetilde{W}}$ & $[-0.47, +0.47]$ & $[-0.98, +0.98]$ & $[-0.83, +0.83]$ \\ 
    $C_{H \widetilde{W} B}$ & $[-0.86, +0.86]$ & $[-1.80, +1.80]$ & $[-1.42, +1.42]$ \\ 
    $C_{H \widetilde{B}}$ & $[-0.74, +0.74]$ & $[-1.55, +1.55]$ & $[-1.32, +1.32]$ \\
    \hline \hline
    \end{tabular*}
\end{table*}

\section{Correlated sensitivities}
\label{sec:2doot}
\begin{figure*}[htb!]
    \centering
    \includegraphics[width=0.19\linewidth]{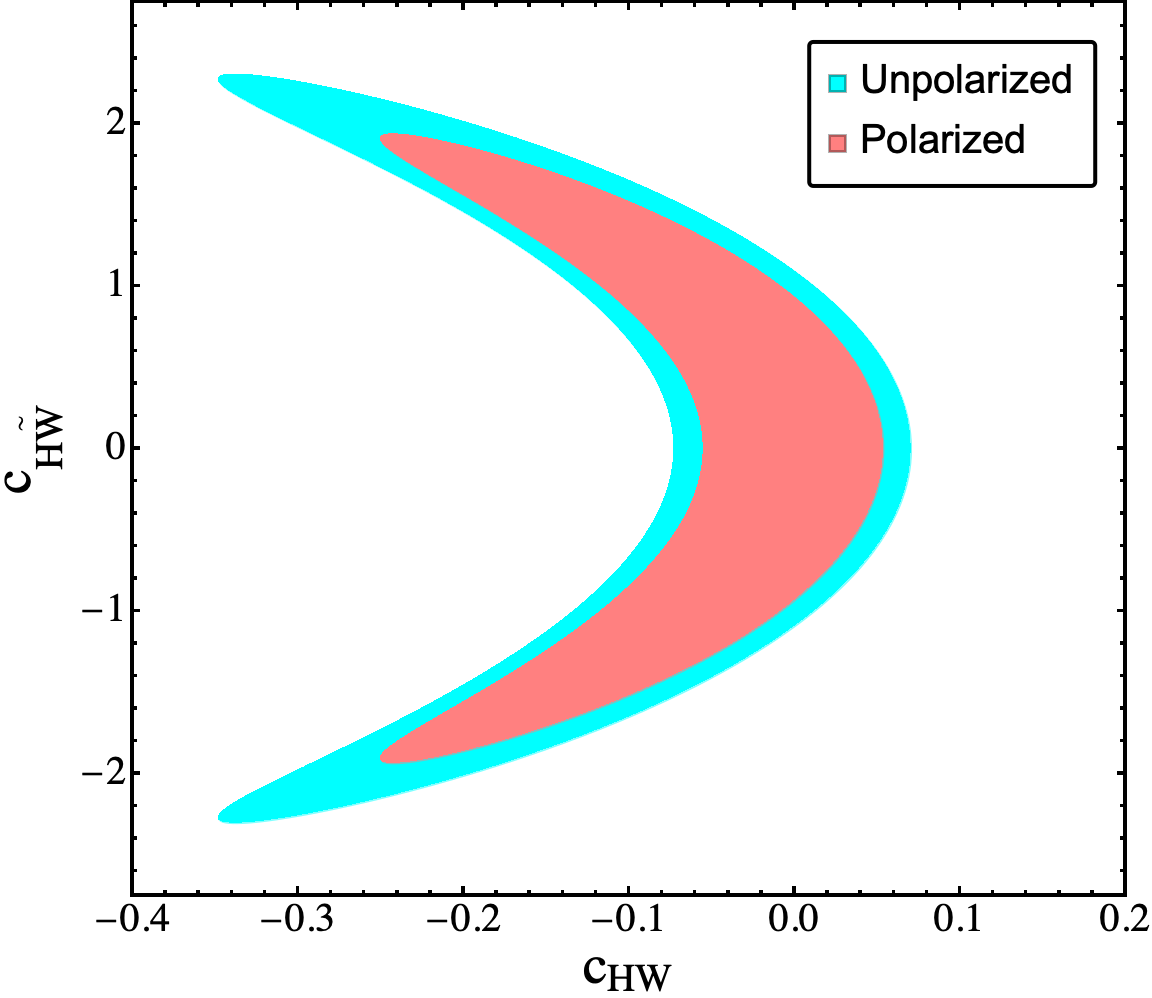}
    \includegraphics[width=0.19\linewidth]{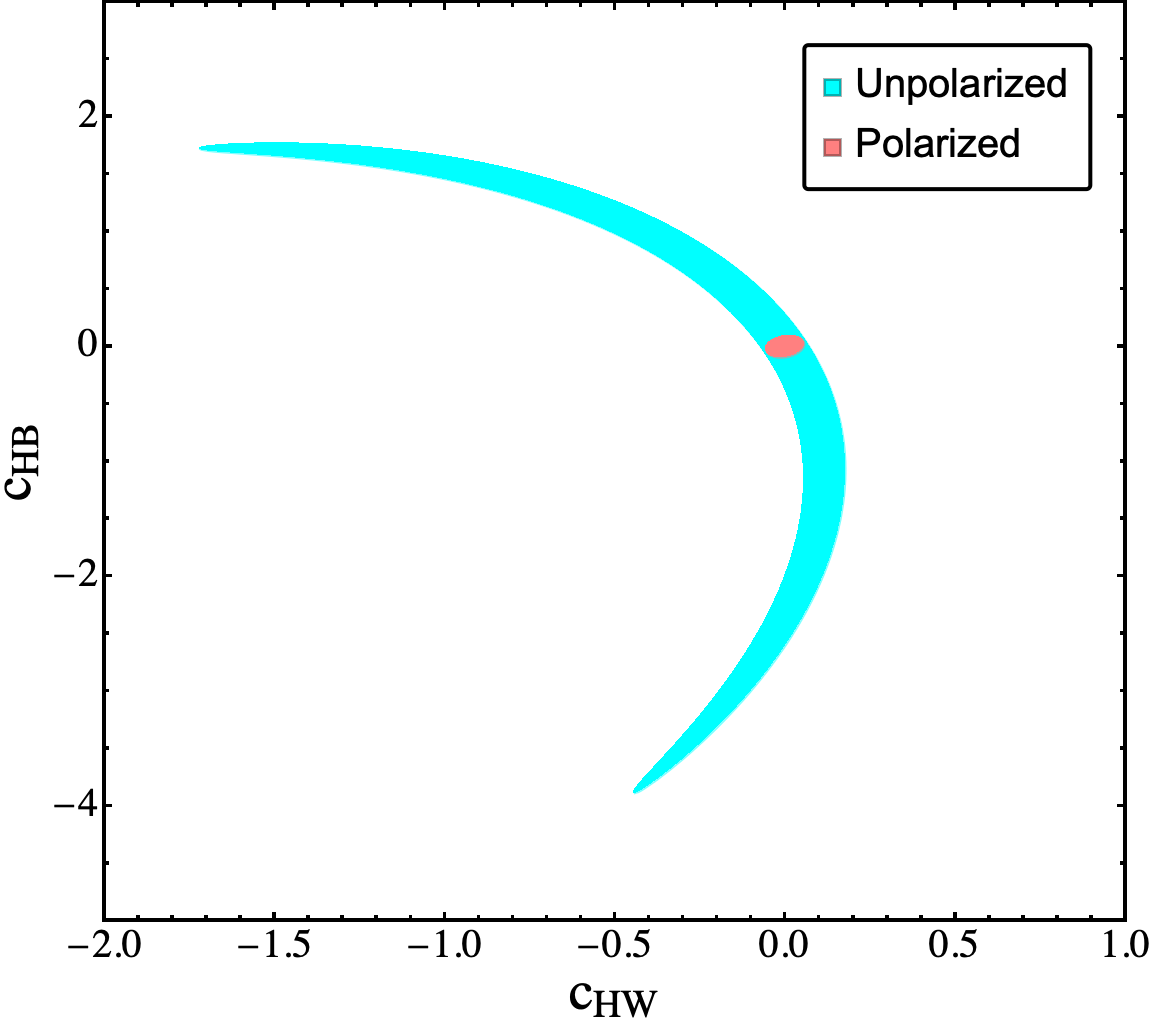}
    \includegraphics[width=0.19\linewidth]{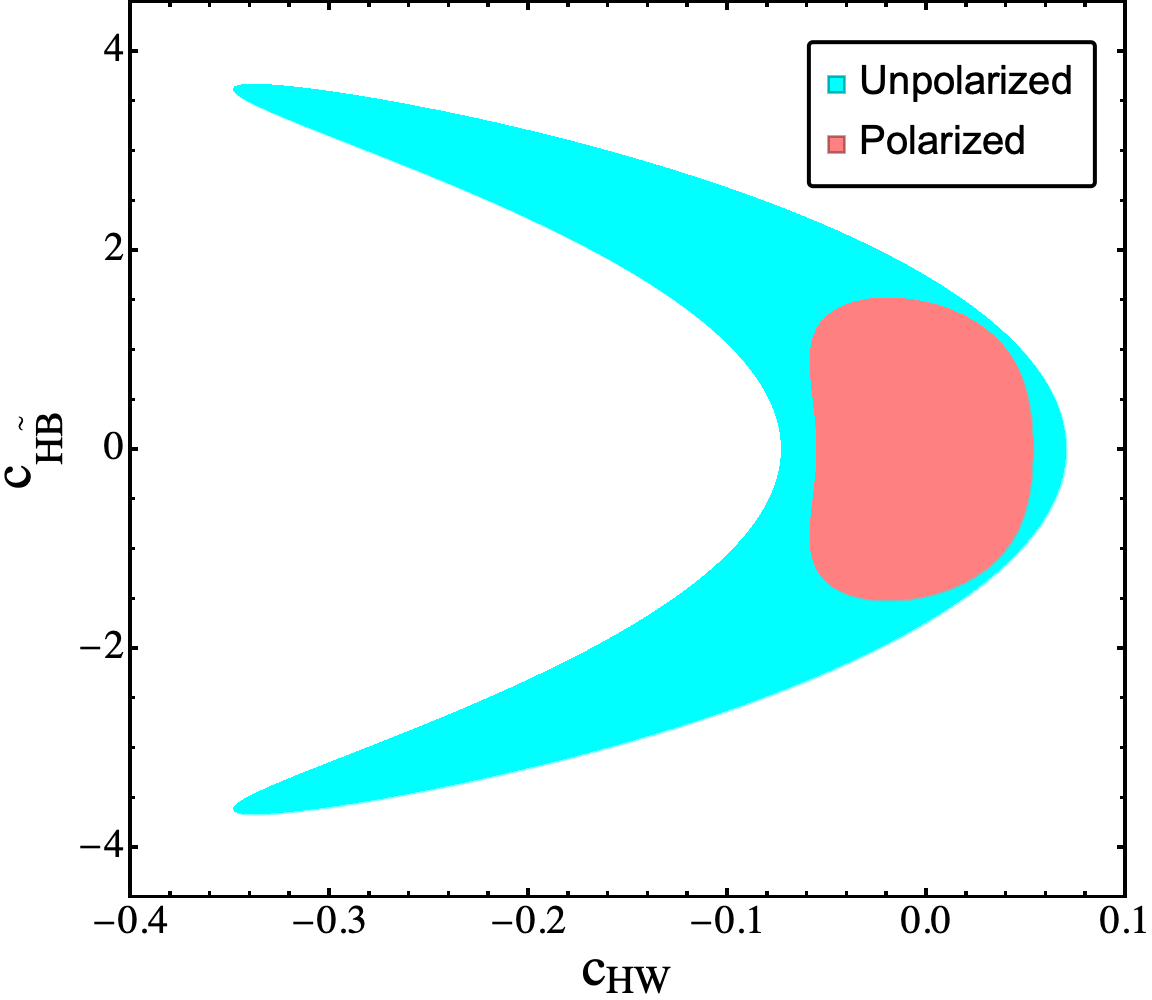}
    \includegraphics[width=0.19\linewidth]{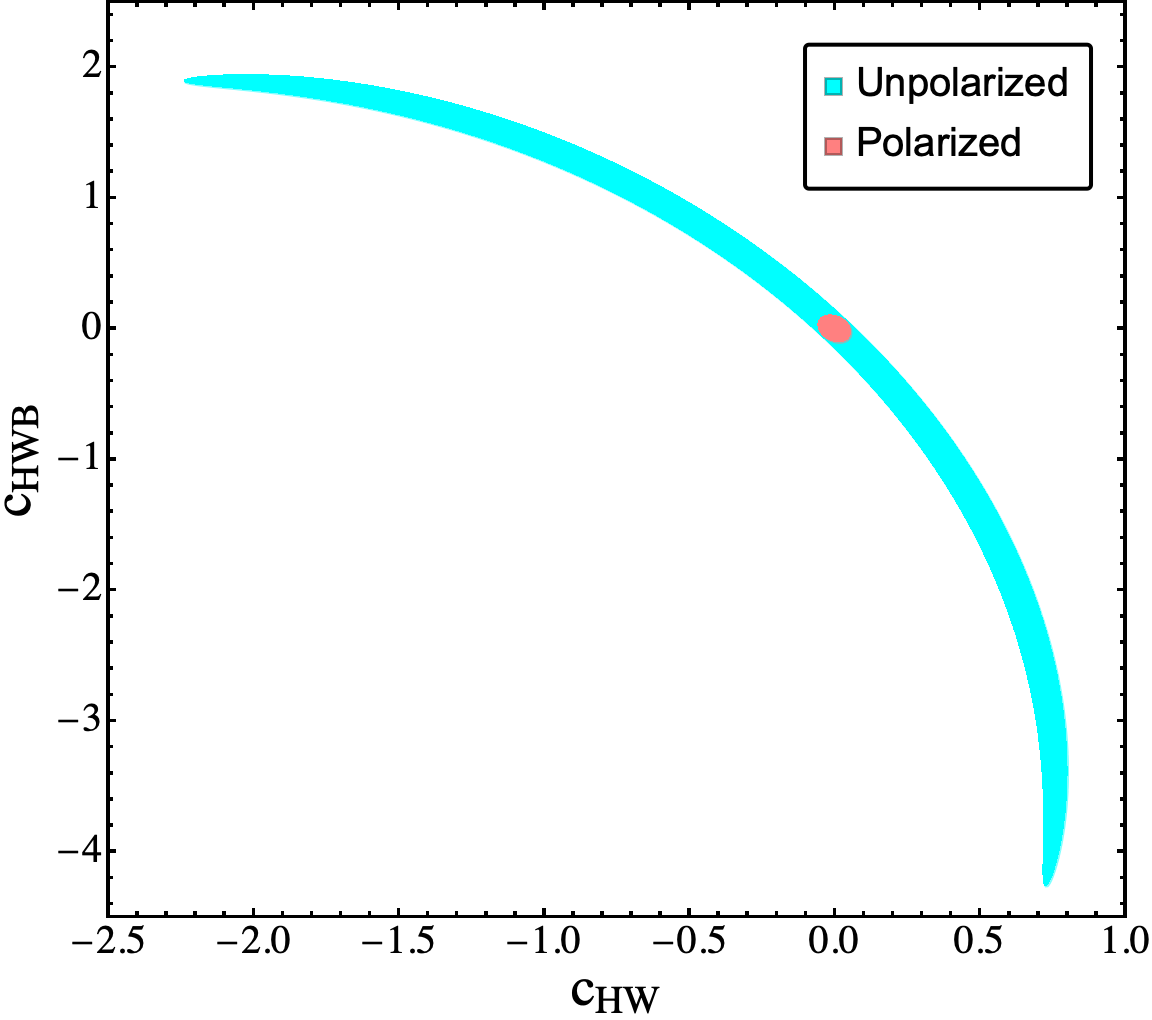}
    \includegraphics[width=0.19\linewidth]{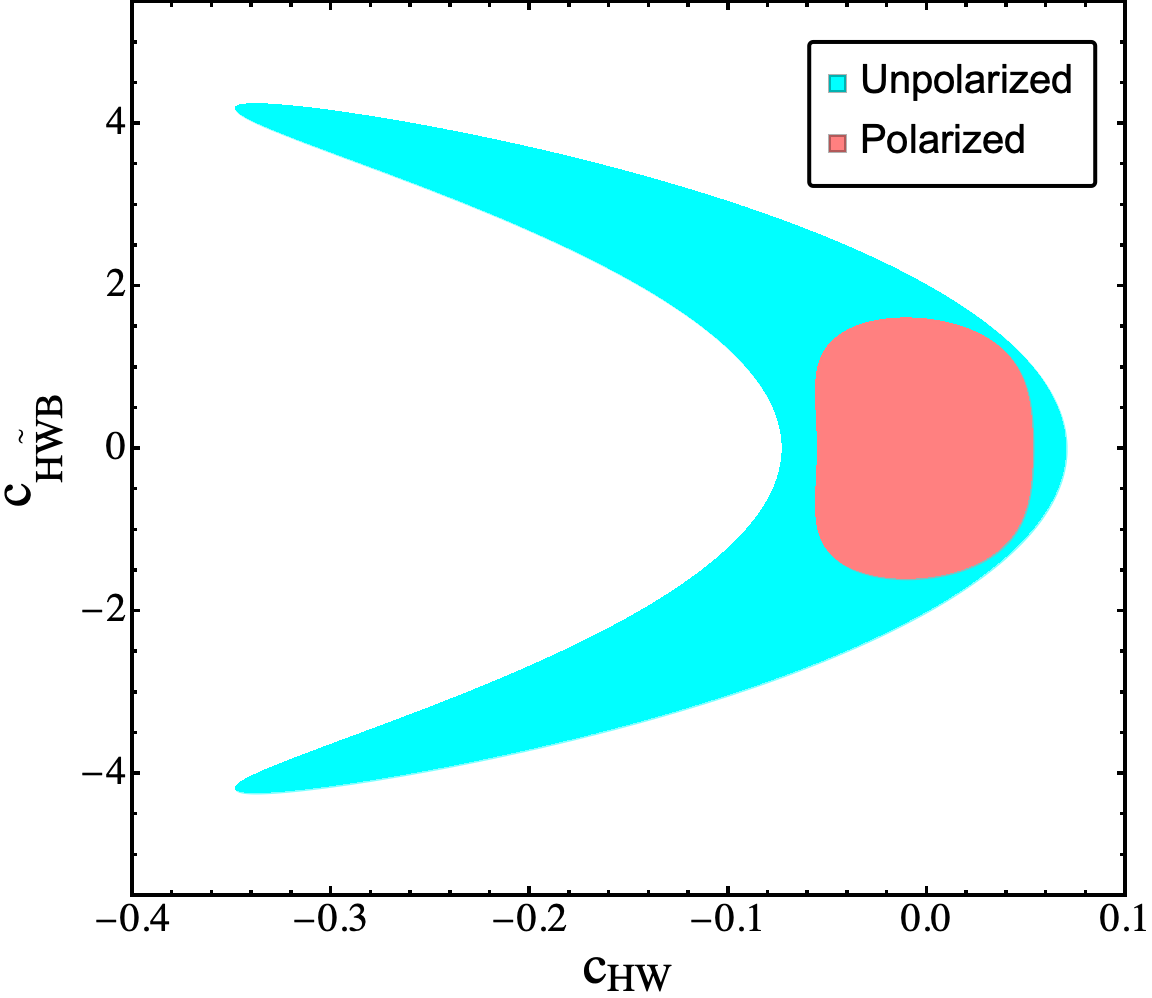}
    \includegraphics[width=0.19\linewidth]{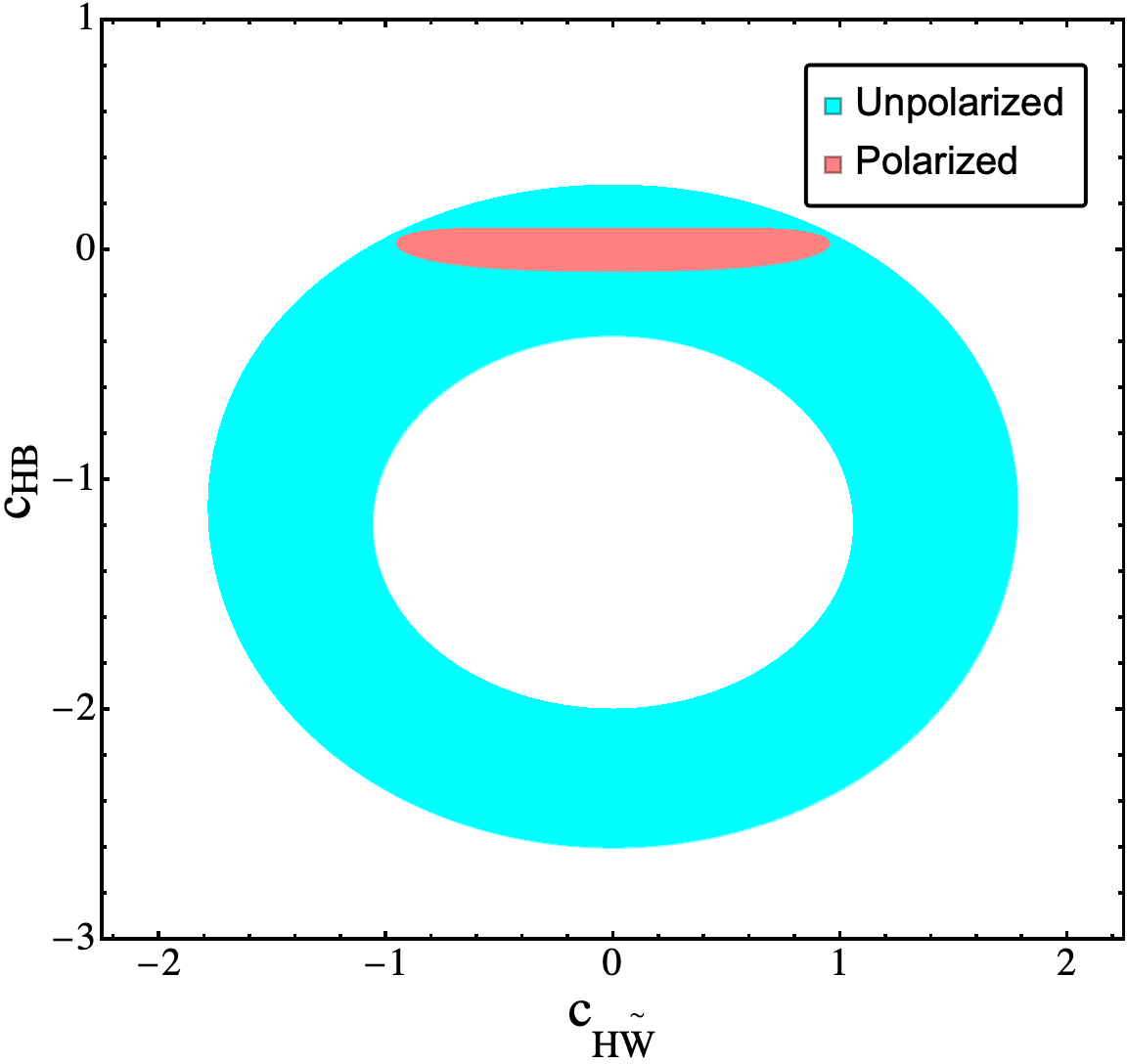}
    \includegraphics[width=0.19\linewidth]{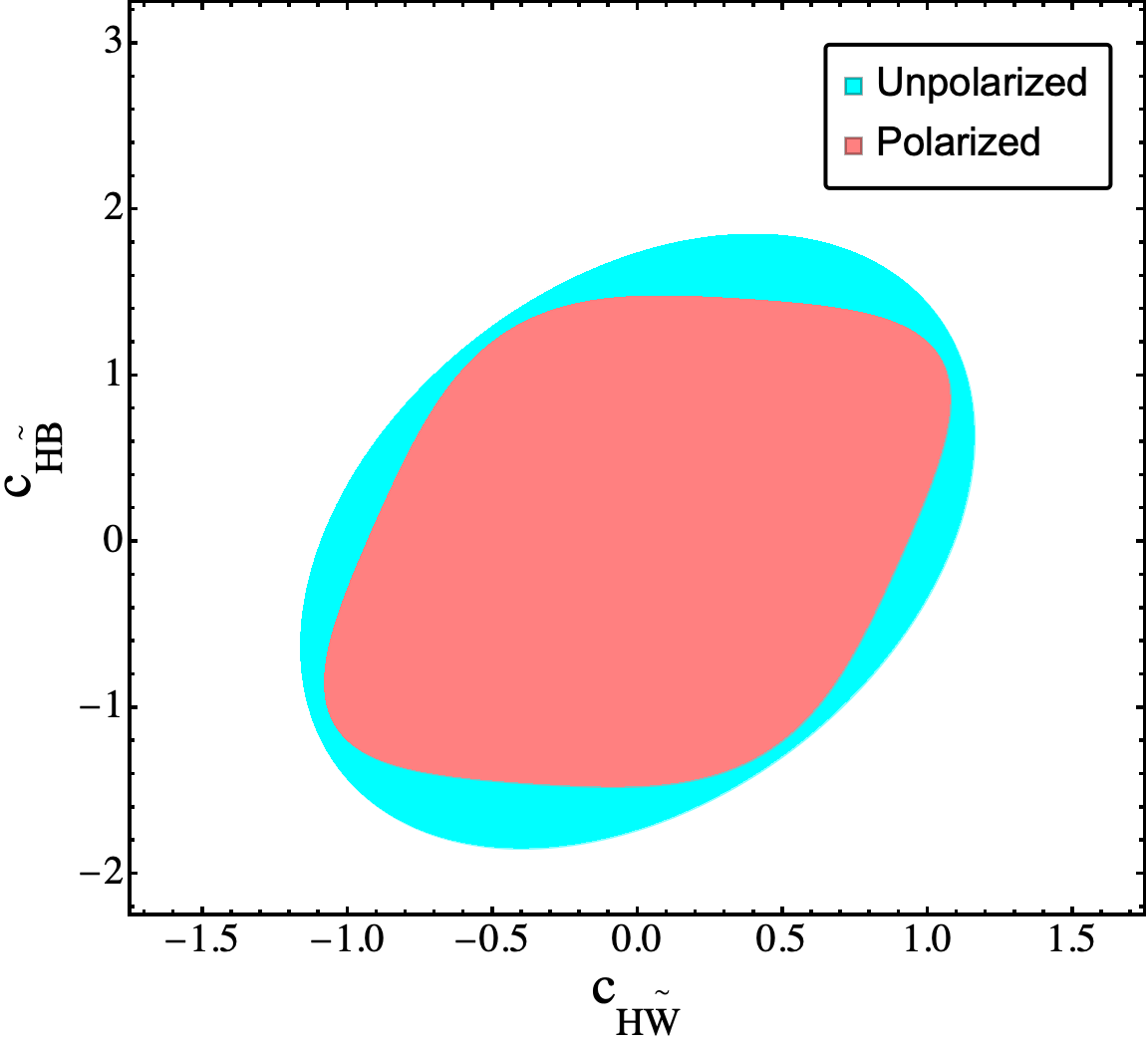}
    \includegraphics[width=0.19\linewidth]{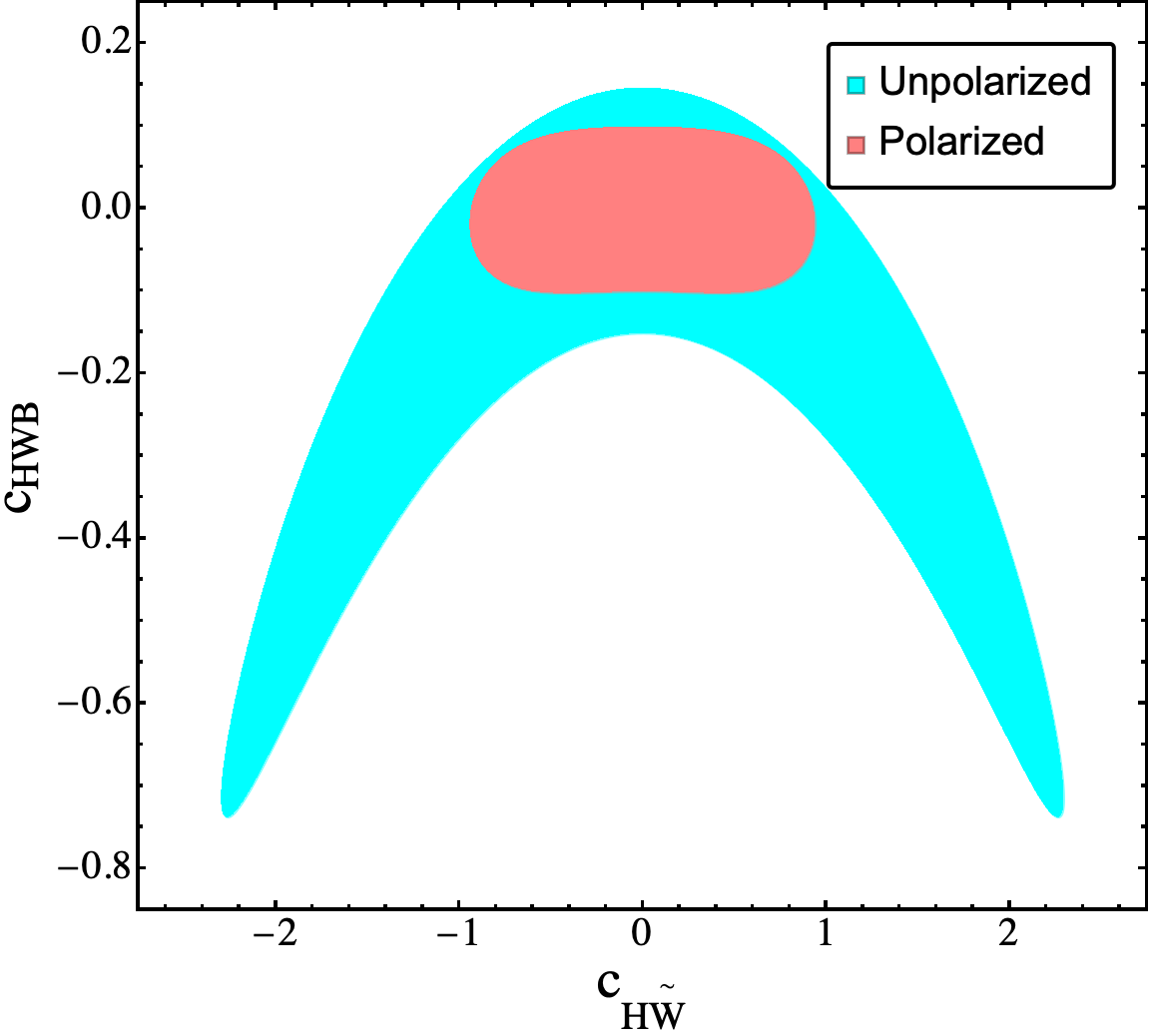}
    \includegraphics[width=0.19\linewidth]{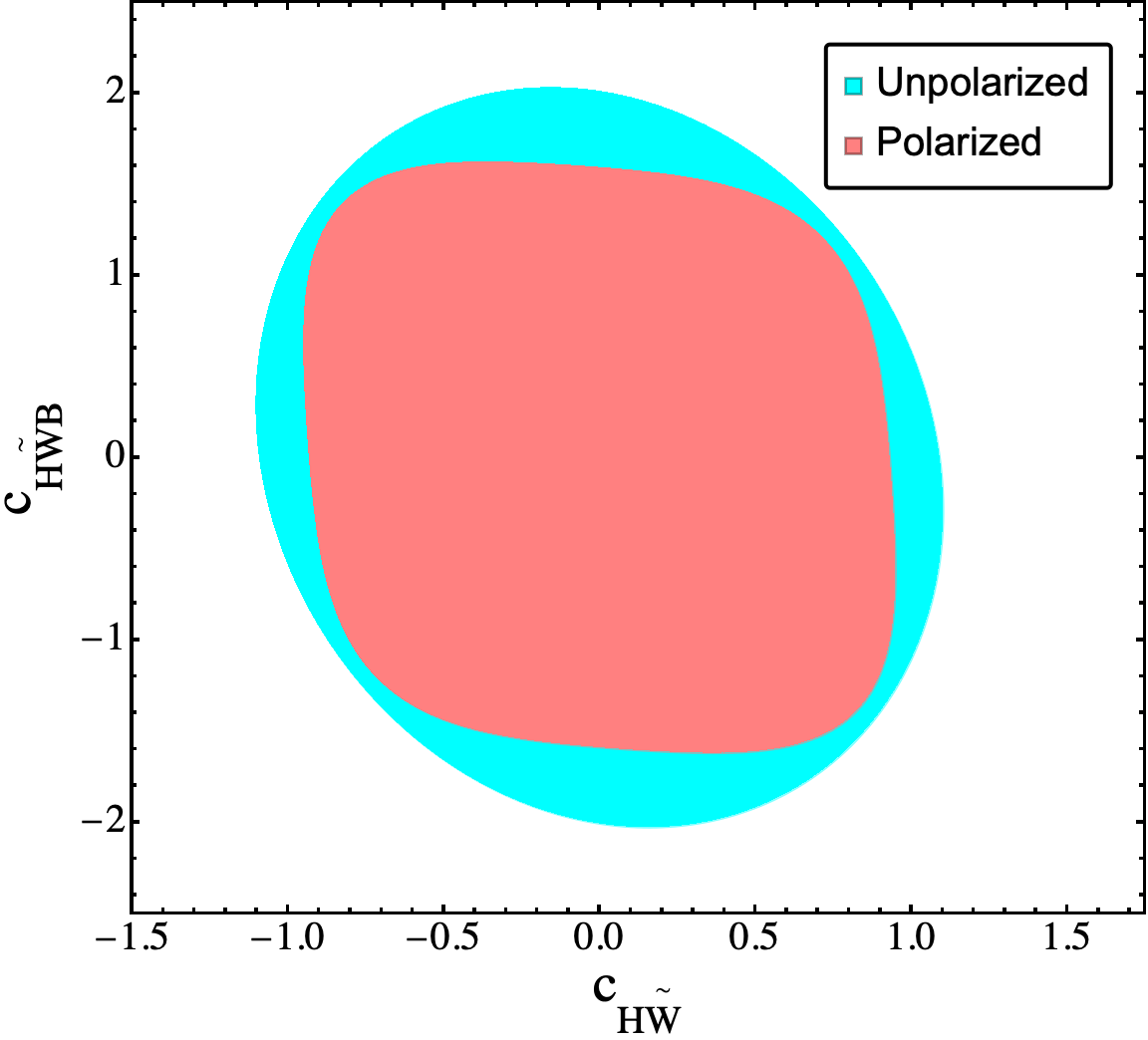}
    \includegraphics[width=0.19\linewidth]{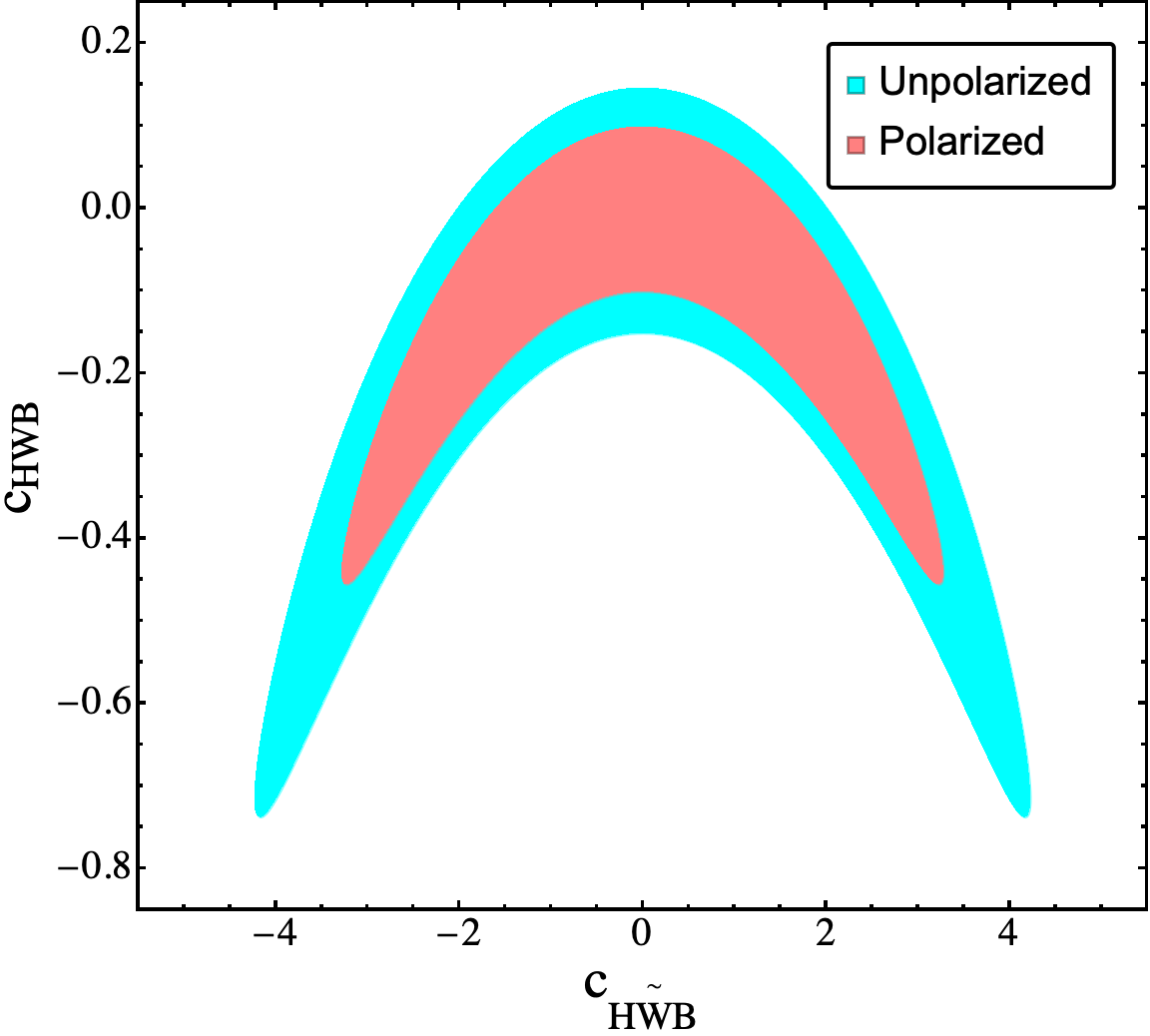}
    \includegraphics[width=0.19\linewidth]{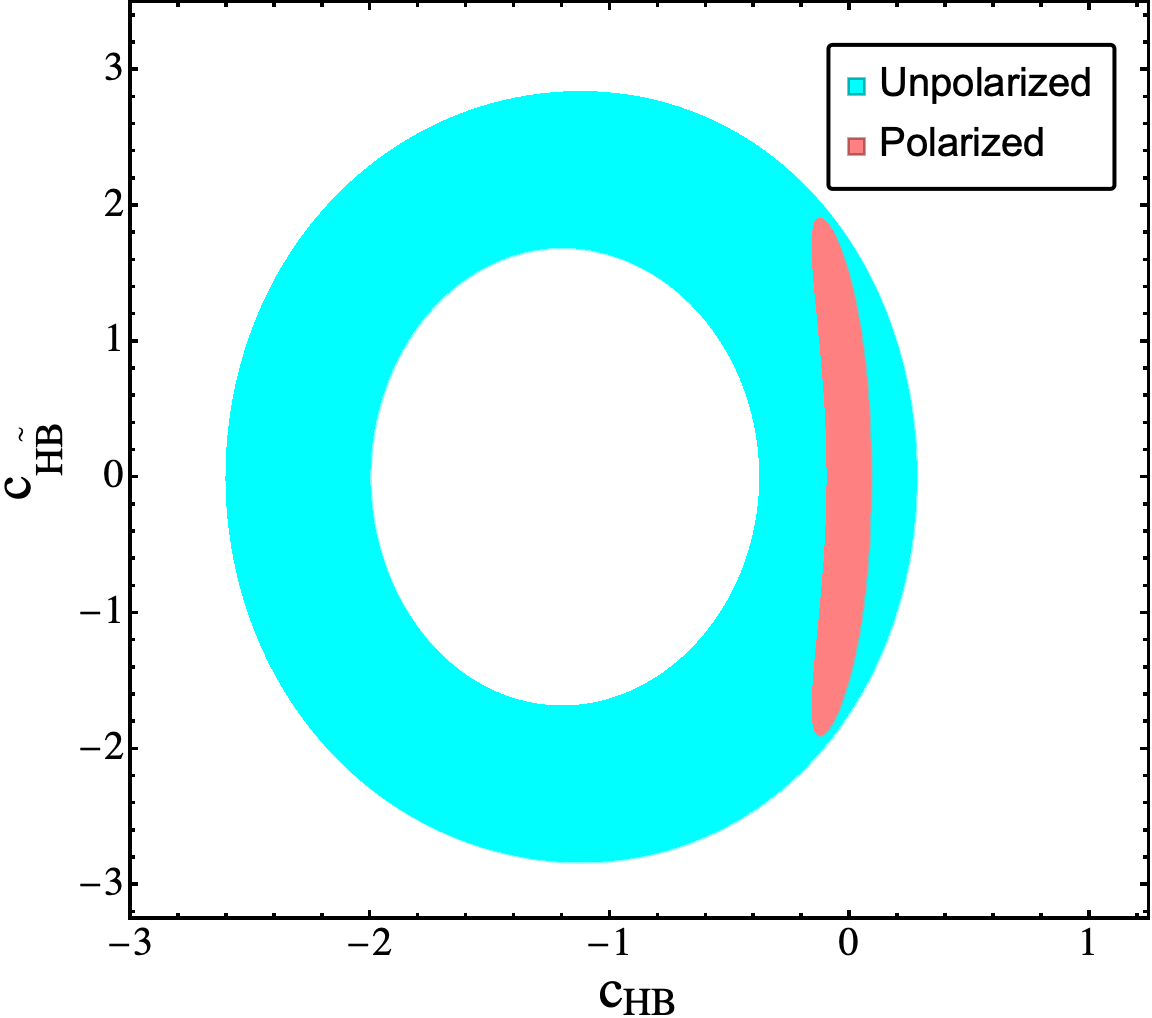}
    \includegraphics[width=0.19\linewidth]{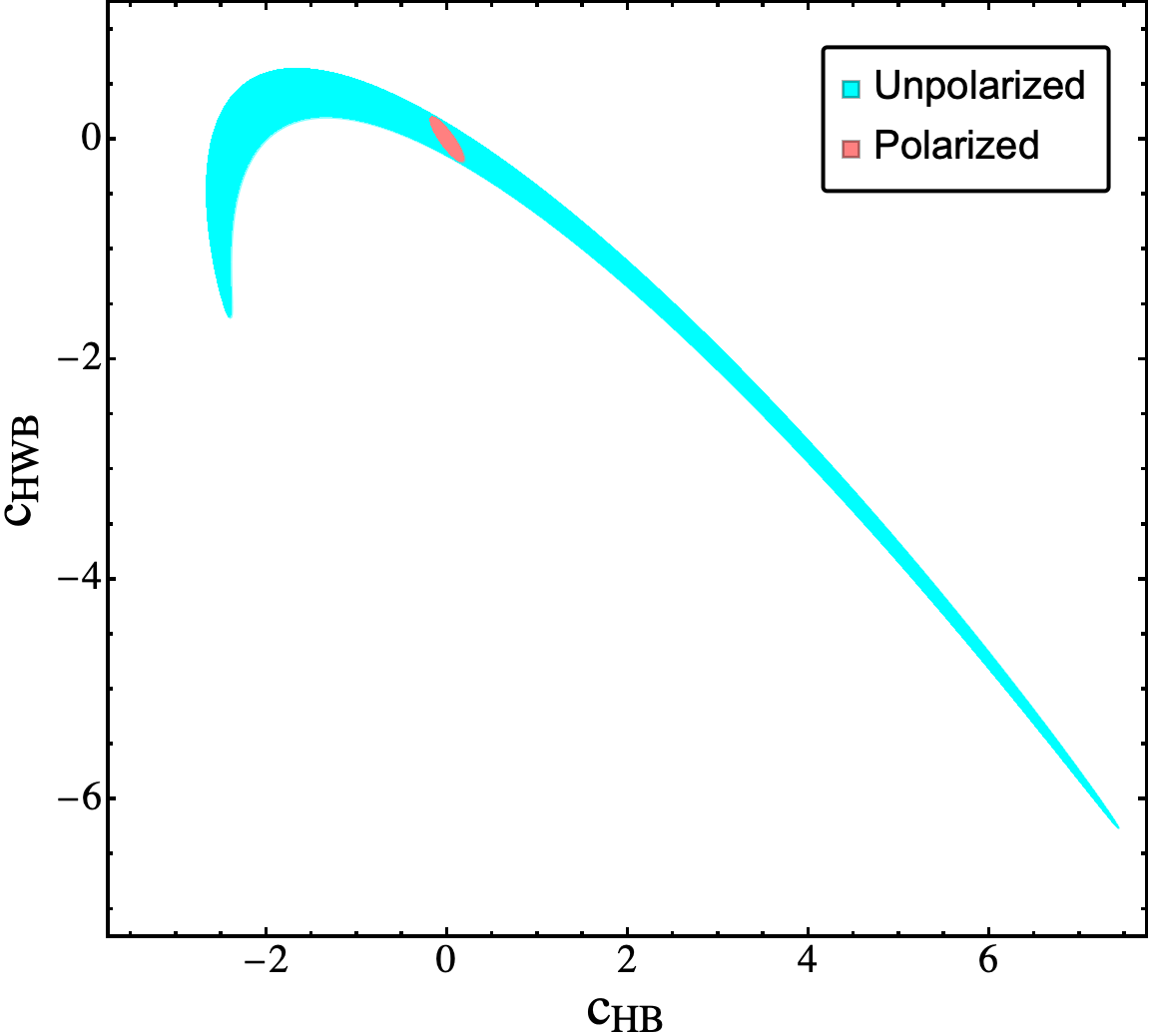}
    \includegraphics[width=0.19\linewidth]{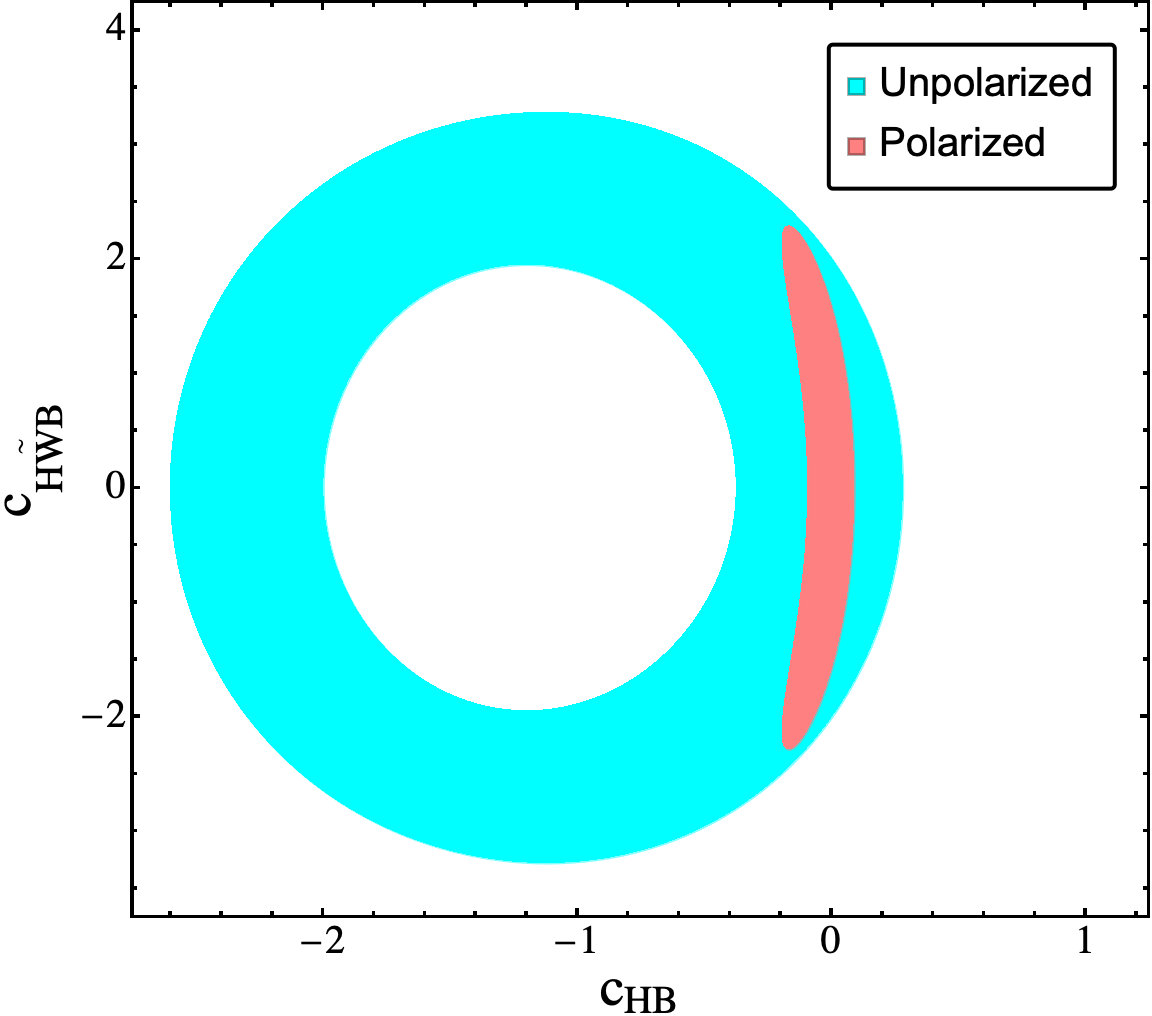}
    \includegraphics[width=0.19\linewidth]{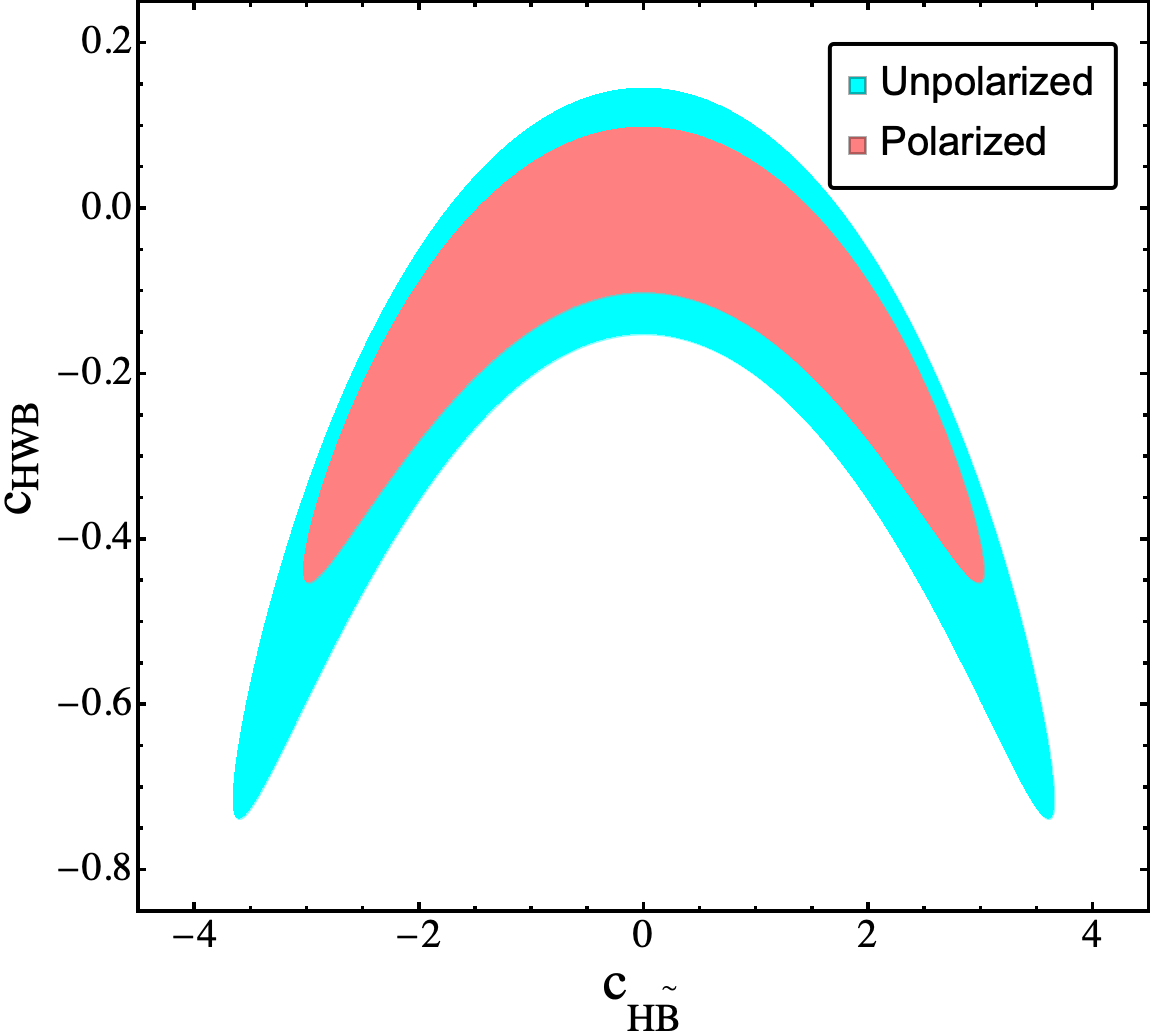}
    \includegraphics[width=0.19\linewidth]{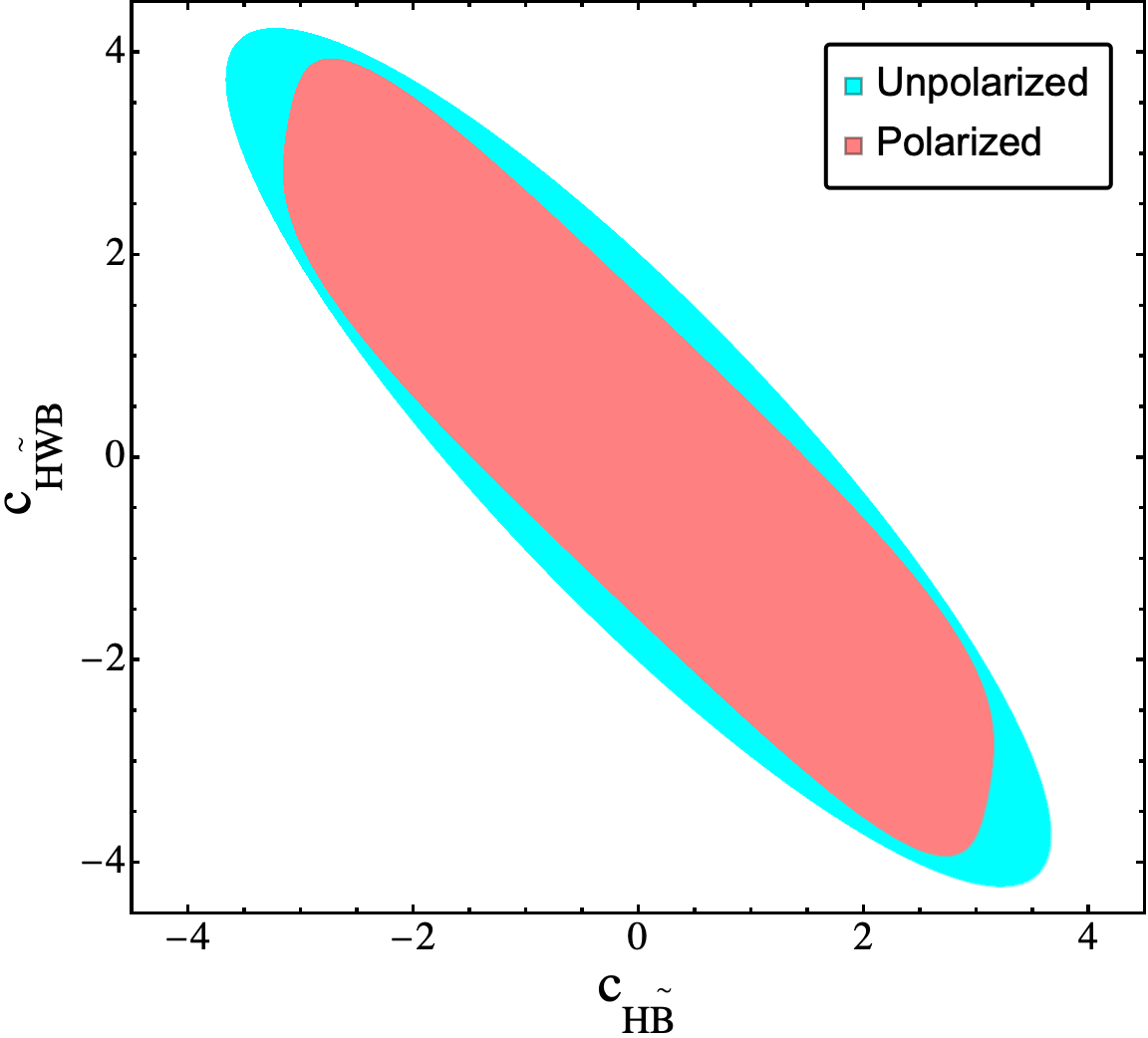}
    \caption{Two parameter optimal sensitivity plots from $Zh$ production at the ILC 250 GeV for unpolarized and combined setup. {\color{blue}For the unpolarized setup, $\mathfrak{L}_{\rm int}=$ 2000 fb$^{-1}$, and for the polarized setup, each polarization configuration viz. $(+30\%,-80\%)$ and $(-30\%,+80\%)$ with $\mathfrak{L}_{\rm int}=$ 1000 fb$^{-1}$, combines to a integrated luminosity of $\mathfrak{L}_{\rm int}=$ 2000 fb$^{-1}$.}}
    \label{fig:oot2d}
\end{figure*}

The two parameter $\chi^2$ sensitivity contours are shown in Fig. \ref{fig:oot2d} for both the unpolarized and the polarized beam configurations discussed in Tab. \ref{tab:oots1}. For the two-parameter case, the 95\% C.L. limit corresponds to $\chi^2 = 5.99$. The comparison between the unpolarized and combined setups clearly demonstrates the enhanced sensitivity achieved through beam polarization, emphasizing its crucial role in constraining SMEFT operators at future $e^+e^-$ colliders.

\bibliography{refer}
\end{document}